\documentclass[11pt]{article}
\usepackage[margin=1in]{geometry}
\usepackage{authblk}
\usepackage[table,dvipsnames]{xcolor}
\definecolor{NCSUred}{RGB}{204,0,0}
\definecolor{NCSUyellow}{RGB}{250,200,0}
\definecolor{NCSUorange}{RGB}{209,73,5}
\definecolor{NCSUdarkred}{RGB}{153,0,0}
\definecolor{NCSUaqua}{RGB}{0,132,115}
\definecolor{NCSUgreen}{RGB}{111,125,28}
\definecolor{NCSUblue}{RGB}{65,86,161}
\definecolor{NCSUlightblue}{RGB}{66,126,147} 
\usepackage{graphicx, epstopdf, subfig, booktabs, cite, url, enumitem}
\usepackage[colorlinks=true, linkcolor=NCSUblue, urlcolor=NCSUaqua, citecolor=NCSUblue]{hyperref}
\usepackage{amsmath, amsfonts, amssymb, amsthm, algorithm2e}
\allowdisplaybreaks
\theoremstyle{plain}
\newtheorem{remark}{Remark}
\newtheorem{assumption}{Assumption}
\newtheorem{definition}{Definition}
\newtheorem{lemma}{Lemma}
\newtheorem{proposition}{Proposition}
\newtheorem{corollary}{Corollary}
\newtheorem{theorem}{Theorem}
\usepackage[capitalize, nameinlink]{cleveref} 

\RequirePackage[T1]{fontenc}
\usepackage{libertinus}
\usepackage{libertinust1math}
\usepackage[cal=rsfso, bb=libus]{mathalpha}

\definecolor{NCSUred}{RGB}{153, 0, 0}
\definecolor{NCSUgreen}{RGB}{0, 132, 115}
\definecolor{NCSUblue}{RGB}{65, 86, 161}
\definecolor{NCSUorange}{RGB}{209, 73, 5}

\newcommand{\mc}[1]{\mathcal{#1}}

\newcommand{\ms}[1]{\mathsf{#1}}
\newcommand{\mr}[1]{\mathrm{#1}}
\newcommand{\mbb}[1]{\mathbb{#1}}

\newcommand{\mR}{\mbb{R}}
\newcommand{\mC}{\mbb{C}}

\newcommand{\mD}{\mbb{D}}

\newcommand{\xD}[1]{\mr{d} #1}
\newcommand{\xDD}[2]{\frac{\xD{#1}}{\xD{#2}}}

\newcommand{\xPP}[2]{\frac{\partial #1}{\partial #2}}
\newcommand{\bra}[1]{\left( #1 \right)}
\newcommand{\Bra}[1]{\left[ #1 \right]}
\newcommand{\BRA}[1]{\left\{ #1 \right\}}
\newcommand{\norm}[1]{\left\| #1 \right\|}
\newcommand{\ip}[2]{\langle #1, \, #2 \rangle}

\newcommand{\vk}{\varkappa}

\title{Data-Driven State Observers for Measure-Preserving Systems
\thanks{This paper was submitted to \textit{IEEE Transactions on Cybernetics} on August 12, 2026. This work is supported by NSF (CBET Award \#2414369). The codes are and data are available at the author's GitHub repository: \url{https://github.com/WentaoTang-Pack/Observer-MeasurePreserving}.}}
\author[1]{Wentao Tang \thanks{Corresponding author: \href{mailto:wtang23@ncsu.edu}{\tt wtang23@ncsu.edu} }}
\affil[1]{Department of Chemical and Biomolecular Engineering, North Carolina State University}
\date{August 12, 2026}

\begin{document}
\pagenumbering{roman}
\maketitle
\pagenumbering{arabic}

\begin{abstract}
The use of data-driven control strategies on systems with not fully measurable states induces the problem of learning-based state observation. 
Motivated by this need, the present work proposes a data-driven approach for the synthesis of state observers for discrete-time nonlinear systems with measure-preserving dynamics. To this end, \emph{Kazantzis--Kravaris/Luenburger (KKL) observers} are shown to be well-defined, where the observer design boils down to determining a nonlinear injective mapping of states and its pseudo-inverse. 
For its learning-based construction, the KKL observer is related to the \emph{Koopman operator}, well-defined on the square-integrable function space and restrictable to a Sobolev-type reproducing kernel Hilbert space (RKHS). 
Hence, observer synthesis algorithms, based on kernel interpolation/regression routines for the desired injective mapping in the observer and its pseudo-inverse, are proposed in various settings of the available dataset -- (i) many orbits, (ii) single long orbit, and (iii) snapshots. Theoretical error analyses are provided, and numerical studies on a chaotic Lorenz system are demonstrated. 
\end{abstract}

\section{Introduction}
The incorporation of machine learning methods in control problems for the development of learning-based control strategies has become a important topic under the name of \emph{data-driven control} \cite{hou2013model, tang2022data}. 
For nonlinear systems, data-driven control often aims at the identification of approximate nonlinear models for model-based controllers \cite{schoukens2019nonlinear}. 
In contrast, in ``direct'' data-driven control, or ``model-free'' control, the modeling procedure becomes optional; instead, data is used to drive the self-adaption of controllers in reinforcement learning \cite{liu2017adaptive, wei2018self, hu2023toward} or provide control-relevant properties (e.g., dissipativity \cite{martin2023guarantees, tang2021dissipativity}). 
It should be noted that in nonlinear control, in order to obtain a state feedback law that guarantees control performance, a \emph{state observer} must be designed to estimate the states from the measurable signals \cite{kravaris2013advances}, whenever the states are not fully measurable. 
Hence, the \emph{data-driven (learning-based) state observer synthesis problem} is intrinsic to the data-driven control theory and practice and thus worth investigation, despite often overlooked in the literature.

\paragraph{Nonlinear observers and their data-driven synthesis}
\par The design of nonlinear state observers, in model-based settings, has been extensively studied; see \cite{bernard2022observer} for a comprehensive review. 
Particularly useful for data-driven synthesis is a generic extended form of Luenberger observer \cite{luenberger1964observing} to nonlinear systems, known as \emph{Kazantzis--Kravaris/Luenberger (KKL) observer} \cite{kazantzis1998nonlinear}, where the observer's internal states follow a linear time-invariant dynamics, with the system outputs as the observer's inputs. 
With this construction, the observer states will ultimately coincide with the system states transformed by a nonlinear injective mapping (henceforth called the \emph{KKL injection}), which satisfies some partial differential equations (\emph{KKL PDEs}) and whose pseudo-inverse will return the estimated states. 
The existence of the KKL injection as well as the admissible choices of the observer parameters have been studied over two decades \cite{andrieu2006existence, bernard2018luenberger, brivadis2023further}, and its data-driven synthesis, being a recent trending topic, mostly follow neural network approaches \cite{ramos2020numerical, niazi2023learning, miao2023learning, tang2024synthesis, woelk2026neural}. 
From a computational point of view, \emph{convex learning}, which guarantees global convergence and efficient solution in a principled way, provides appealing alternatives (e.g., online regression \cite{tang2023data} and kernel canonical correlation analysis \cite{woelk2025data}). We therefore adopt this convex learning methodology here. 

In the author's recent works, noting the relation between the KKL injection and the \emph{Koopman operator} or \emph{Koopman semigroup}, the idea of \emph{Koopman-based KKL observer} has been examined. 
In \cite{ye2026}, for nonlinear dynamics near the origin, by approximating the Koopman operator in finite dimensions, the error is accounted for as uncertainties in a robust formulation of observer synthesis. 
In \cite{ni2026}, for limit-cycle dynamics, Koopman eigenfunctions associated with given eigenvalues are estimated, which guarantee a rich basis for the regression of KKL injection.  
Conceptually, the Koopman operator, which maps any state-dependent function (on a properly defined function space) to its composition with the system's transition or flow mapping, ``lifts'' nonlinear dynamics to a linear system in the infinite-dimensional function space \cite{brunton2022modern}. Hence, for KKL observers, where the nonlinear observation problem is resolved by finding a \emph{nonlinear} injection satisfying a \emph{linear} observer dynamics, the Koopman framework is relevant as expected.

\paragraph{Measure-preserving systems} Measure-preserving dynamics are common in realistic nonlinear systems, where the states evolve with a nontrivial invariant measure instead of converging to an equilibrium point. 
In addition to systems with limit cycle or limit torus dynamics, where the invariant measure is supported on a low-dimensional manifold, we are also particularly interested in \emph{chaotic dynamics}, hallmarked by unbounded Lyapunov exponents and fractional-dimensional strange attractors \cite{strogatz2024nonlinear}.  
Chaos often exist in the gait patterns and path planning behaviors of robots \cite{zang2016applications, ban2022Dynamic}, electric power systems \cite{oneill-carrillo1999chaotic, yu2003power}, and chemical reaction kinetics in manufacturing systems \cite{teymour1989dynamic, balakotaiah1999bifurcation}. 

For data-driven analysis of measure-preserving systems, one then encounters the challenge of a generic treatment with the resulting Koopman operators, which may possess a continuous spectrum without any eigenvalue or eigenfunction \cite{mezic2020spectrum}. 
A precise characterization of the Koopman spectrum thus entails the learning of its \emph{spectral resolution} or \emph{spectral measure}. 
In \cite{korda2020data}, based on ergodic sampling on the attractor, the autocorrelation and its Fourier transform are estimated to reconstruct the spectral measure. 
In \cite{colbrook2024rigorous} and \cite{boulle2025convergent}, in view of possible spectral pollution (spurious approximate eigenfunctions/eigenvalues), residual-based dynamic mode decomposition (ResDMD) was proposed for \textit{a posteriori} verification spectral properties. 

In this paper, we focus on the problem of \emph{data-driven state observer synthesis for measure-preserving nonlinear systems}. 
Since the observer obtained in the existing model-based \cite{solak2001observer, hua2004adaptive} or data-driven ways (e.g. \cite{tang2024synthesis, woelk2026neural}) is agnostic of the intrinsic behavior of measure-preserving flows, the resulting observation performance can be restrictive. 
By using the Koopman operator with the knowledge of the measure preservation, the data-driven observer synthesis is thus more structured and capable of exploiting Koopman spectrum. 

\paragraph{Contributions and organization of this work}
The present paper have the following technical contributions.
\begin{enumerate}
    \item A discrete-time KKL observer is proposed, where the KKL injection is defined on the entire attractor with proven injectivity under appropriate backward distinguishability conditions. 
    \item With the KKL injection related to the Koopman operator, we adopt a two-stage learning-based approach, comprising of the data-based construction of the KKL injection and the regression of its pseudo-inverse. These learning routines are based on kernel methods. 
    \item The synthesis of such a KKL observer is considered in three different settings of available dataset -- (i) many orbits, each containing the history backtracked from a sample point, (ii) one long orbit that covers the attractor very densely, and (iii) snapshots of predecessor--successor states. Building on the existing theories of kernel interpolation and kernel regression, the error bounds for these synthesis approaches are given. 
\end{enumerate}
The proposed Koopman-based KKL observer is applied to a Lorenz system with comparison to a model-based Luenberger observer, exhibiting high observation accuracy despite severe nonlinearity (which causes extended Kalman filter to fail). 
The readers will note that the proposed approach is not only data-driven (not using the governing equations), but also computational efficient --- \emph{no nonconvex optimization is needed for training, and the performance is theoretically guaranteed}, which sharply contrasts all neural network approaches. 

\par The remainder of this paper is organized as follows. In \S\ref{sec:prelim}, preliminaries on KKL observers, Koopman operators, and RKHSs are introduced. 
In \S\ref{sec:existence}, the Koopman spectrum structure is examined, the injectivity of a KKL observer is proved, and its connection to the Koopman operator is shown. 
The learning-based constructions of the KKL observer under various data settings, as well as the resulting error bounds, are discussed in \S\ref{sec:computation}. 
The numerical experiments on the Lorenz system are shown in \S\ref{sec:numerical}, followed by conclusions in \S\ref{sec:conclusions}. 

\paragraph{Notation} We use lower-case letters for scalar- and vector-valued objects, and upper-case letters for matrix- and operator-valued objects. 
Euclidean spaces and their subsets use blackboard letters, e.g., $\mbb{X}\subset \mR^d$, $\mbb{Z}_{a,b} = \mbb{Z}\cap [a, b]$, and $\mbb{D} = \{z\in \mbb{C}: |z|\leq 1\}$. 
$\partial^\alpha$ is the partial derivative with multi-index $\alpha \in \mbb{N}^d$; the length of $\alpha = (\alpha_1, \cdots, \alpha_d)$ is $|\alpha|=\alpha_1+\cdots + \alpha_d$. 
We denote by $\mu^\circ$ the standard Lebesgue measure. 
Function spaces use calligraphic letters, e.g., $\mc{N}$ (reproducing kernel Hilbert space) and $\mc{H}$ (Sobolev--Hilbert space). The composition operation between functions are denoted as $\circ$, namely $g\circ f: x\mapsto g(f(x))$. Inner products are denoted as $\ip{\cdot}{\cdot}$. 
The adjoint operator of an operator $K$ is denoted by $K^\ast$; consistently the complex conjugate of $z\in \mC$ is $z^\ast$. Finally, $\mr{i} = \sqrt{-1}$ and $\mr{C}_n^k = n!/((n-k)!k!)$.

\section{Preliminaries}\label{sec:prelim}
Here we consider a discrete-time system represented as
\begin{equation}\label{eq:system}
x_{t+1} = f(x_t), \quad y_t = h(x_t), \quad t\in \mbb{Z}. 
\end{equation}
where $f: \mbb{X} \subset \mR^{d_x} \rightarrow \mbb{X}$ and $h: \mbb{X}\rightarrow \mR$ are continuous. 
As a standing assumption, $\mbb{X}$ is a bounded closed region with a boundary of zero Lebesgue measure and a regular interior.\footnote{
The need for this regularity arises from the need to identify Sobolev spaces with an RKHS (to be discussed in \S\ref{subsec:RKHS}). By a \emph{regular} open set $\mbb{O} \subset \mR^{d_x}$, we mean that (i) the boundary $\partial \mbb{O}$ is a smooth manifold, and near any point, $\mbb{O}$ is on one side of $\partial\mbb{O}$, (ii) there exists a sufficiently smooth mapping $\psi: \mR^{d_x-1}\rightarrow \mR$, such that at any $x_0\in \partial \mbb{O}$, $\partial_{d_x} \psi(x_0) \neq 0$, (iii) $\forall x_0\in \partial \mbb{O}$, there exists a neighborhood $\mbb{U}_0\ni x_0$, such that upon the coordinate transform $\bar{\psi}(\xi) = (\xi_1, \dots, \xi_{d_x-1}, \psi(\xi_1,\dots,\xi_{d_x-1}))$, satisfies: $\bar{\psi}(\mbb{O}\cup \mbb{U}_0) \subset \{\xi\in \mR^{d_x}: \xi_{d_x}>0\}$ and $\bar{\psi}(\partial\mbb{O}\cup \mbb{U}_0) \subset \{\xi\in \mR^d: \xi_{d_x}=0\}$.}
According to the Krylov--Bogolyubov theorem, there exists an invariant, finite, Borel measure $\mu$ on the state space $\mbb{X}$; that is, $\mu(\mbb{A}) = \mu(f^{-1}(\mbb{A}))$ for any Borel subset $\mbb{A}$ of $\mbb{X}$. 
Generally, the invariant measure may not be supported over the entire $\mbb{X}$ but only the attractor. For example, if the system is attracted to a single point, then the invariant measure supported on a singleton. We are interested in the cases where the attractor is non-trivial and observe the dynamics on the attractor. Hence, we consider tacitly $\mbb{X}$ as the attractor set.

\subsection{KKL observer}
Let us start with a formal definition of a state observer for \eqref{eq:system}. Conceptually, a state observer is a dynamical system by itself, and hence has its own states, here referred to as \emph{observer states} and denoted as $z$, whose dynamics contains the system's measurable outputs $y$ as the inputs, i.e.,
\begin{equation}\label{eq:KKL.state}
    z_{t+1} = f_z(z_t, y_t), \quad t\in \mbb{Z}. 
\end{equation}
In order that the signal $z$ contains ``all the information'' of the system states $x$, we require that there is an injective (one-to-one) mapping $\zeta: \mbb{X}\rightarrow \zeta(\mbb{X})$, such that if $\zeta(x_0) = z_0$, then $\zeta(x_t) = z_t$ holds for all $t\geq 0$. 
As such, the injection $\zeta$ has a left pseudo-inverse, denoted as $\zeta^\dagger$, such that $\zeta^\dagger \circ \zeta = \mr{id}_\mbb{X}: \mbb{X} \rightarrow \mbb{X}$, $x\mapsto x$. Hence, we let the state estimate be
\begin{equation}\label{eq:KKL.output}
    \hat{x}_t = \zeta^\dagger(z_t), \quad t\in \mbb{Z}. 
\end{equation}
If $\zeta(x_0) = z_0$, then $\hat{x}_t$ coincides with the actual state $x_t$. Clearly, such an injection $\zeta$, if exists, must satisfy:
\begin{equation}\label{eq:functional.equation}
    \zeta(f(x)) = f_z(\zeta(x), h(x)), \quad \forall x\in \mbb{X}. 
\end{equation}
Thus, the observer synthesis problem entails three aspects: 
\begin{enumerate}[label=(\roman*)]
    \item designing the observer state dynamics $f_z$,
    \item solving $\zeta$ that satisfies functional equation \eqref{eq:functional.equation}, and 
    \item retrieving its pseudo-inverse mapping. 
\end{enumerate} 

\par The KKL observer resolves the first issue by assigning a linear time-invariant (LTI) state dynamics to the observer.
\begin{definition}
    A \emph{Kazantzis--Kravaris/Luenberger (KKL) observer} is one specified by \eqref{eq:KKL.state} and \eqref{eq:KKL.output}, with $f_z(z, y) = Az + by$. 
\end{definition}
Thus, the observer synthesis problem reduces to the determination of the KKL injection $\zeta$, which must solve a linear functional equation:
\begin{equation}\label{eq:KKL.FE}
    \zeta\circ f = A\zeta + bh. 
\end{equation}
In \cite{kazantzis1998nonlinear}, the conditions for the existence of such an injection solution for \eqref{eq:KKL.FE} \emph{near an equilibrium point} (say, the origin $0$ in $\mR^{d_x}$) include the following ones:
\begin{enumerate}
    \item \emph{Hyperbolicity.} The locally linearized dynamics of $f$, namely the Jacobian, $F=\mr{D}f(0)$, does not have any eigenvalue on the unit circle $\partial \mbb{D}$. 
    \item \emph{Local observability.} The observability matrix $[H; HF; \dots; HF^{d_x -1}]$ of the linearized system at the origin has rank $d_x$, where $H = \mr{D}h(0)$. 
    \item \emph{Observer controllability.} The controllability matrix $[b, Ab, \dots, A^{n-1}b]$ of the observer has rank $d_x$. 
    \item \emph{Non-resonance.} For any eigenvalue of $A$, $\lambda_j(A)$, there do not exist nonnegative integers $k_1, \dots, k_{d_x}$ such that $\sum_{i=1}^n k_i > 0$ and $\prod_{i=1}^n \lambda_i(F)^{k_i} = \lambda_j(A)$. 
\end{enumerate}
In addition to the existence of an injective $\zeta$, it is desired that $z_t - \zeta(x_t)\rightarrow 0$ as $t\rightarrow 0$ even if the initialization of the observer is not exact ($z_0 - \zeta(x_0) \neq 0$). For this, we need $A$ to be \emph{stable}, i.e., all eigenvalues of $A$ need to be inside $\mbb{D}$. 
Although these conditions are mild, the \emph{locality} essentially restricts the use of KKL observers near an equilibrium point, almost exclusively applicable to self-stabilizing dynamics. 

\par In \cite{andrieu2006existence} and a series of works thereafter \cite{bernard2018luenberger, brivadis2019luenberger, brivadis2023further, tran2023arbitrarily}, the conditions for KKL observers to exist on the entire state region $\mbb{X}$ are studied.
For the discrete-time system \eqref{eq:system}, the conditions in \cite{brivadis2019luenberger} are as follows:
\begin{enumerate}[label=(\roman*)]
    \item \emph{Invertibility and Lipschitzness.} $f$ is invertible, with a $\mc{C}^1$ (continuously differentiable) inverse, on $\mR^{d_x}$, and $h\in \mc{C}^1(\mR^{d_x})$. Moreover, $f^{-1}$ and $h$ are globally Lipschitz. 
    \item \emph{Backward distinguishability.} For all $x\neq x'$ in $\mbb{X}$, there exists a time $-\ell<0$ such that $h(f^{-\ell}(x))\neq h(f^{-\ell}(x'))$. 
\end{enumerate}
These conditions guarantee that $(A,b)$ can be chosen almost arbitrarily as any sufficiently fast \footnote{According to \cite{brivadis2019luenberger}, the eigenvalues of $A$ must reside in $c_1^{-1}\mbb{D}$ whenever $c_1 > 1$, where $c_1$ refers to the constant such that $\|x\|\leq c_0 + c_1\|f(x)\|$ with another constant $c_0$. Such a choice may be conservative, which we actually can avoid in this work.} 
stable and controllable pair, if the dimension of $z$ is set as ${d_x}+1$.

To compute the KKL injection, an explicit formula can be found, if further assuming that $f^{-1}$ is valued on $\mbb{X}$ (namely that the dynamics is \emph{backward invariant} on $\mbb{X}$) and $h$ is bounded on $\mbb{X}$:
\begin{equation}\label{eq:Brivadis.series}
 \zeta(x) = \sum_{\ell=0}^\infty A^\ell b h\bra{f^{-(\ell+1)} (x)}. 
\end{equation}
In fact, only in this case can we find an analytical expression for $\zeta$ and also guarantee its injectivity. 
Practically, the local solution for $\zeta$ and its left pseudo-inverse $\zeta^\dagger$ may follow a formal power series approach \cite{kazantzis1998nonlinear, kazantzis2001discrete}. In principle, any other numerical methods, including the learning methods mentioned in the Introduction, can also be used to solve \eqref{eq:KKL.FE}. 
Examining the KKL functional equation \eqref{eq:KKL.FE}, it is not hard to find that the equation is a \emph{linear functional equation}, although involving nonlinear dynamics $f$. 
Hence, it is naturally connected to the well-known concept of Koopman operator.

\subsection{Koopman operator} 
\begin{definition}[\cite{brunton2022modern}]
    The \emph{Koopman operator} for system \eqref{eq:system} is the following linear operator:
    \begin{equation}
        K: \quad \mc{G}\rightarrow \mc{G}, \quad g \mapsto g\circ f.  
    \end{equation}
    The domain $\mc{G}$ can be any space of scalar-valued functions on $\mbb{X}$ that guarantees $g\circ f\in \mc{G}$, $\forall g\in \mc{G}$. Any member of $\mc{G}$ is called an ``observable'' function. 
\end{definition}
To choose a function space $\mc{G}$ guaranteeing the invariance under composition, a typical choice is $\mc{G} = \mc{L}^2_\mu$, the collection of functions $g$ such that $\|g\|_{\mc L^2_\mu}^2 :=\int_{\mbb{X}} |g|^2 d\mu < \infty$. 
Indeed, it can be easily verified, due to the invariance of $\mu$, that the Koopman operator $K$ is a bounded operator and moreover an \emph{isometry}: $\|Kg\|_{\mc{L}^2_\mu}^2 = \|g\|_{\mc{L}^2_\mu}^2$.
However, for computation, $\mc{L}^2_\mu$ may be inconvenient due to the need of sampling under the (unknown) invariant measure $\mu$ and choosing a basis of such a Hilbert space. Also, the functions on $\mc{L}^2_\mu$ are not pointwise defined and hence may not be continuous, which makes them difficult to recover from data.  

\par 
In view of these limitations of $\mc{L}^2_\mu$, we instead consider a Sobolev--Hilbert space $\mc{H}^s(\mbb{X})$, comprising of every function $g$ that has generalized derivatives\footnote{By a generalized derivative of $g \in \mc{L}^2$ over $x_i$, we refer to a function $u\in \mc{L}^2$ satisfying $\int_{\mbb{X}} u\varphi \xD{x} = -\int_{\mbb{X}} g\xPP{\varphi}{x_i} \xD{x}$ for any $\varphi \in \mc{C}^\infty(\mbb{X})$ with a compact support in $\mbb{X}$. We still denote $u=\xPP{g}{x_i}$. Recursively, for any multi-index $\alpha=(\alpha_1, \dots,\alpha_d)$, higher-order generalized derivatives $\partial^{\alpha_1 +\cdots+ \alpha_d} g / \partial x_1^{\alpha_1}\cdots \partial x_d^{\alpha_d}$ are defined and simply denoted as $\partial^\alpha g$.} up to the $s$-th order (where $s$ is a positive integer), all of which being $\mc{L}^2$ with respect to the Lebesgue measure \cite{adams2003sobolev}. 
On $\mc{H}^s(\mbb{X})$, the squared norm of a function $g$ is defined as
$$ \|g\|_{\mc{H}^s(\mbb{X})}^2 =  \sum_{0\leq |\alpha|\leq s} \| \partial^\alpha g\|_{\mc{L}^2(\mbb{X}) }^{2}, $$
where $\alpha$ ranges over all the multi-indices with length $|\alpha| = \alpha_1+\dots+\alpha_d\leq s$. 
We shall first see that if the dynamics is regular enough, then the Koopman operator can be well-defined on a Sobolev--Hilbert space without involving the invariant measure. 
\begin{proposition}[Koopman operator on $\mc H^s$ \cite{kohne2025error}]
\label{prop:Koopman.boundedness}
    If $f\in [\mc{C}^s(\mbb{X})]^{d_x}$, $\mbb{X}\subset \mR^{d_x}$ is bounded, and $\inf_{x\in \mbb{X}} |\det \mr{D}f(x)|>0$, then $K: \mc{H}^s(\mbb{X}) \rightarrow \mc{H}^s(\mbb{X})$ is a bounded linear operator. 
\end{proposition}

We should also examine the \emph{inverse} of the Koopman operator, since observation is intrinsically a backward-in-time task. 
\begin{assumption}\label{assum:backward}
    $f\in [\mc{C}^s(\mbb{X})]^{d_x}$ is invertible, and $f^{-1}\in [\mc{C}^s(\mbb{X})]^{d_x}$. In addition $\inf_{x\in \mbb{X}} |\det \mr{D}f(x)|$ and $\inf_{x\in \mbb{X}} |\det \mr{D}f^{-1}(x)|$ are both positive. 
\end{assumption}
Under the assumption, the dynamics on $\mbb{X}$ is forward and backward invariant. As the backward dynamics is at least as regular as the forward dynamics, then it naturally follows that the backward Koopman operator $K^{-1}$, is also bounded as a linear operator on $\mc{H}^s(\mbb{X})$. Obviously, $K^{-1}: g\mapsto g\circ f^{-1}$. 
\begin{corollary}
    Under Assumption \ref{assum:backward}, both $K$ and $K^{-1}$ are linear bounded operators on $\mc{H}^s(\mbb{X})$. \footnote{
    The index $s$ of the Sobolev--Hilbert space on which the Koopman operator is defined matches smoothness of the dynamics $f$. Hence, the higher smoothness we know of $f$, the smaller space we can restrict the Koopman operator to. In other words, we choose a function space that is just ``as large as needed''. 
    Typically, if the system is governed by physically meaningful terms, the smoothness $s$ is sufficiently high and often known \textit{a priori}, even without detailed equations.
    In particular, if the system is continuous-time with all constituent terms in $\mc{C}^s$, then the flow is $\mc{C}^s$. 
    With a small sampling interval $\Delta t$, the conditions on the non-degeneracy of Jacobians can be easily met, since $\mr{D}f(x)\to I$ as $\Delta t\to0$. 
    }
\end{corollary}
\begin{remark}\label{rem:fractal}
    It is possible to have a fractal attractor set $\mbb{X}$. In this case, the integral concepts involved in the definitions of generalized derivative and Hilbert--Sobolev spaces cannot be defined based on the usual concept of Lebesgue measure. Instead, Hausdorff measure should be used as a generalization of the standard Lebesgue measure, cf. the textbook of \cite{stein2009real}, Chapter 7. 
\end{remark}

Next, in the machine learning context, we introduce an equivalent representation of $\mc{H}^s(\mbb{X})$ as a reproducing kernel Hilbert space (RKHS). With this, $K$ and $K^{-1}$ remain to be bounded linear operators in such a Sobolev-type RKHS.

\subsection{Learning in reproducing kernel Hilbert spaces}\label{subsec:RKHS}
\begin{definition}
    A continuous bivariate function $\vk: \mbb{X}\times\mbb{X}\rightarrow \mR$ is called a \emph{Mercer kernel}, if for any finite number of points: $x_1, \dots, x_n\in \mbb{X}$, the matrix formed by $\vk(x_i, x_j)$ ($i,j\in \mbb{Z}_{1,n}$), called the kernel matrix, is positive semidefinite. A reproducing kernel Hilbert spaces (RKHS), or native space, refers to the completion of $\mr{span}\{\vk(x, \cdot):x\in \mbb{X}\}$, conferred with the inner product: $\ip{\vk(x, \cdot)}{\vk(x', \cdot)} = \vk(x, x')$. We denote the RKHS with kernel $\vk$ on $\mbb{X}$ as $\mc{N}_\vk(\mbb{X})$. 
\end{definition}

As an implication of the inner product definition on RKHS, we know that for any $f\in \mc{N}_\vk(\mbb X)$, $$\ip{f}{\vk(x, \cdot)} = f(x) ; $$
this is known as the \emph{reproducing property} \cite{steinwart2008support}. With this nice property, RKHSs have been found useful in statistical learning problems (colloquially known as the kernel method) \cite{smola1998learning}, such as kernel regression, kernel support vector machine, and kernel principal component analysis. 
The learning of Koopman operator or its adjoint operator on RKHSs was been first proposed in \cite{kevrekidis2016kernel} through a kernel version of the dynamic mode decomposition (DMD) algorithm and then as regression problems of Hilbert--Schmidt operators \cite{klus2020eigendecompositions, kostic2022learning}. 
From the previous discussions, we see that $K$ and $K^{-1}$ are bounded linear operators on $\mc{H}^s(\mbb{X})$ provided regularity assumptions. 
To formulate learning problems regarding the Koopman operator, it is therefore desirable to equate $\mc{H}^s(\mbb{X})$ with some RKHS --- such an idea was initially proposed in \cite{kohne2025error} and commonly used in recent works \cite{boulle2025convergent, tang2025koopman, tang-ye2025koopman}. 
We provide the proposition below without proof, but only point out to the readers that essentially it uses a characterization of the Sobolev space by the Fourier transform $\hat{g}$ of $g$: $$\mc{H}^s(\mR^{d_x}) = \BRA{g: \mbb{X}\rightarrow \mR: \int_{\mR^{d_x}} (1+\|\xi\|^2)^s |\hat{g}(\xi)|^2 \xD{\xi} < \infty }. $$

\begin{proposition}[Sobolev-type RKHS \cite{wendland2004scattered}]
\label{prop:radial.kernel}
    Suppose that $\mbb{X}\subset \mR^{d_x}$ is a bounded region with regular boundary, and $\vk$ is a radial kernel, i.e., there exists a function $\rho: [0, \infty) \rightarrow \mR$ such that $\vk(x, x') = \rho(\|x-x'\|)$ for all $x, x'\in \mR^{d_x}$. If the Fourier transform $\hat{\rho}$ satisfies
    $$c_1(1+\|\xi\|^2)^{-s} \leq \hat{\rho}(\xi) \leq c_2(1+\|\xi\|^2)^{-s} $$
    for some constants $0< c_1 < c_2 < \infty$, and $s>d_x/2$, then $\mc{N}_\vk(\mbb{X}) \equiv \mc{H}^s(\mbb{X})$ with equivalent norms. 
\end{proposition}
\begin{remark}
    The condition that $s>d_x/2$ is necessary, as only in this case, the Sobolev space $\mc{H}^s(\mR^d)$ can be continuously embedded in a (H{\"{o}}lder-)continuous function space and thus a RKHS defined by a continuous kernel. In the case of $s=0$ (namely $\mc{H}^0 = \mc{L}^2$), the condition on Fourier transform cannot be satisfied, justifying the undesirability of using $\mc{L}^2$ space for learning. 
\end{remark}

To satisfy the condition in Proposition \ref{prop:radial.kernel}, we use the following kernels \cite{wendland2004scattered}, called the \emph{Wendland kernels}: 
\begin{equation}\label{eq:Wendland.kernel}
\begin{aligned}
    &\vk_{d_x,k}^{\ms{W}}(x,x') = \rho_{d_x, k}^{\ms{W}}(\|x-x'\|), \, 
    \rho_{d_x,k}^{\ms{W}} = I^k \varrho_{\lfloor d_x/2+k+1 \rfloor}^{\ms{W}} \\ 
    &\text{where } \varrho_\ell^{\ms{W}}(r) = \max\{1-r, 0\}^\ell, \, I\phi(r) = \int_r^\infty r'\phi(r') \mr{d}r'. 
\end{aligned}
\end{equation}
When using $\vk=\vk_{d_x, k}^{\ms{W}}$ ($k=0,1,2,\cdots$), $\mc{N}_\vk(\mR^d) \equiv \mc{H}^{s}(\mR^d)$ with $s=\frac{d_x+1}{2} + k$. Of course, this equivalence remains if the distance between $x$ and $x'$ is rescaled by a constant $\sigma>0$. 
Another possible choice is the \emph{Mat{\'{e}}rn kernel}:
\begin{equation}\label{eq:Matern.kernel}
    \vk_{d_x, k}^{\sf{M}}(x,x') = \frac{\bra{\sqrt{2k} \|x-x'\|}^k}{\Gamma(k)2^{k-1}}  \mr{K}_k \bra{\sqrt{2k} \|x-x'\|},
\end{equation}
in which $\mr{K}_k$ ($k=0,1,2,\cdots$) is the $k$-th modified Bessel function of the second type and $\Gamma$ is the Gamma function. When the Mat{\'{e}}rn kernel $\vk=\vk_{d_x, k}^{\ms{M}}$ is adopted, $\mc{N}_\vk(\mbb{X})$ is equivalent to $\mc{H}^{s}(\mbb{X})$ with $s=d_x/2 + k$.

\begin{remark} 
    If the resulting $s$ is fractional, the Sobolev--Hilbert space is still well-defined. In this case, it is an interpolation space between $s_1=\lfloor s\rfloor$ and $s_2 = \lceil s\rceil$. As long as $K$ is well-defined as a linear bounded operator on $\mc{H}^{s_1}(\mbb{X})$ and $\mc{H}^{s_2}(\mbb{X})$, by Marcinkiewicz interpolation theorem, $K$ is a linear bounded operator on $\mc{N}_\vk(\mbb{X})$. Then the following discussions will still apply. 
\end{remark}

\begin{remark}\label{rem:submanifold}
    The assumption that $\mbb{X}$ is a compact set with regular interior is technically necessary for the continuous extension from $\mc{H}^s(\mbb{X})$ to $\mc{H}^s(\mR^{d_x})$ entailed in the proof of Proposition \ref{prop:radial.kernel} \cite{wendland2004scattered}. This assumption, however, can be violated in two situations -- (i) when $\mbb{X}$ is only a compact submanifold of $\mR^{d_x}$, and (ii) when $\mbb{X}$ is fractal. 
    \begin{itemize}
        \item In the first case, assuming sufficient smoothness of the submanifold, there exist a finite number of bounded open subsets covering $\mbb{X}$, each smoothly diffeomorphic to a bounded open subset of $\mbb{R}^{d_x'}$ (where $d_x'\leq d_x$ is a positive integer) by a coordinate chart. Hence, substituting $d_x$ by $d_x'$ in the above proposition still leads to a valid conclusion and there is no hampering on later discussions. 
        \item In the second case, the equivalence between the Sobolev space and RKHS becomes obscure; in fact, the Sobolev space cannot be defined based on the usual concept of derivatives. We can only take the conclusion of Proposition \ref{prop:radial.kernel} as an \emph{ad hoc} assumption. 
        It may still be conjectured that the conclusion holds true by replacing $d_x$ with the Hausdorff dimension; however, the author is not aware of such discussions in the literature. 
    \end{itemize}
\end{remark}

Combining the previous two propositions, we conclude that once the dynamical system $f$ is known to be sufficiently smooth, then there is an RKHS on which the Koopman operator $K$ and its inverse are bounded linear operators. The calculation with $K$ on an RKHS is brought by its adjoint. 
\begin{definition}
    The \emph{adjoint operator} of $K$ is the operator $K^\ast: \mc{N}_\vk(\mbb{X}) \rightarrow \mc{N}_\vk(\mbb{X})$ such that for all $g_1, g_2\in \mc{N}_\vk(\mbb{X})$, $\ip{K^\ast g_1}{g_2} = \ip{g_1}{Kg_2}$. It can be verified that:
    \begin{equation}\label{eq:Perron-Frobenius}
        K^\ast \vk(x, \cdot) = \vk(f(x), \cdot), \quad \forall x\in \mbb{X}. 
    \end{equation}
\end{definition} 
An interesting property that follows is that if $\vk$ is designed as such that $\vk(x,x) = 1$ for all $x\in \mbb{X}$ (as is the case with Wendland kernels), we always have $\|K^\ast \vk(x, \cdot)\| = \|\vk(x, \cdot)\|$. This, however, does not imply that $K$ is an isometry on the RKHS, since the norm may not be preserved on the span of kernel functions. Instead, $\|K\|\geq 1$ should always be true.

\section{Existence of KKL Observer for Measure-Preserving Systems}\label{sec:existence} 
Now we consider whether a KKL observer is indeed well-defined. First, we briefly discuss the structure of the Koopman operator's spectrum. 
Then, we discuss the existence of a KKL observer, consistent with the ``deep KKL observer'' design in \cite{brivadis2023further} for continuous-time systems. Finally, we express the KKL observer in terms of the Koopman operator. 

\subsection{Koopman spectrum}
The understanding of Koopman spectrum is of interest to the prediction of functions under the dynamics. 
By spectrum of a linear bounded operator $K$, denoted as $\mbb{Sp}(K)$, we refer to the set of complex numbers $\lambda \in \mC$ such that $K-\lambda$ does not have a bounded inverse. 
Intuitively, if $\psi\in \mc{G}$ is an eigenfunction of $K$ with eigenvalue $\lambda$ (if it exists), then $\psi\circ f^t(x) = K^t\psi(x) = \lambda^t\psi(x)$. 
Since the observer is conceptually related to the backward dynamics $f^{-1}$, we are naturally also concerned with the spectrum of the backward Koopman operator $K^{-1}$. 
First of all, in our setting (under Assumption \ref{assum:backward}), $K$ can not be compact. If $K$ is a compact operator, the structure of its spectrum is close to that of a finite-rank operator. That is, there are only point spectrum (eigenvalues) and continuous spectrum (the set of points $\lambda\in \mC$ such that the range of $K-\lambda$ is not the entire $\mc{G}$ but dense in $\mc{G}$). Moreover, there can be at most countably many nonzero spectral points, each being an eigenvalue with a finite-dimensional eigen-space, and their only possible cluster point is $0$. Hence, a compact operator can be approximated by a finite-rank operator to any precision in operator norm. 
However, the compactness of $K$ would imply that $KK^{-1} = \mr{id}$ (the identity map on $\mc{G}$) is compact, which is obviously false. 

\par In the case of $\mc{G} = \mc{L}^2_\mu$, it is guaranteed that $K$ is an isometry, and so is $K^{-1}$. Thus, $K$ has a spectral resolution:
\begin{equation}\label{eq:spectral.resolution}
    K = \int_{\partial\mbb{D}} \lambda \xD{E}(\lambda). 
\end{equation}
Here, $E$ is an \emph{operator-valued measure}, called the \emph{spectral family} or \emph{resolution of identity}, of $K$. That is, given any Borel measurable set $\mbb{A} \in \partial\mbb{D}$, it is ensured that $E(\mbb{A})$ is a projection on $\mc{L}^2_\mu$, with $E(\partial\mbb{D}) = \mr{id}$, and $E$ is countably additive with respect to its arguments. 
Thus, we have $$ K^{-1} = \int_{\partial\mbb{D}} \lambda^{-1} \xD{E}(\lambda) \text{ and } K^\ast = \int_{\partial\mbb{D}} \lambda^\ast \xD{E}(\lambda). $$ To perform any operation on these operators, we simply need the information of the spectral family $E$, which can be estimated from ergodic sampling data \cite{korda2020optimal}. 
We are interested in the situation when $\mc{L}_\mu^2(\mbb{X}) \equiv \mc{L}^2(\mbb{X})$, which can be guaranteed by the assumption below. 
\begin{assumption}\label{assum:measure-preserving}
    The invariant measure $\mu$ and Lebesgue measure $\mu^\circ$ are absolutely continuous mutually. That is, the Radon--Nikodym density $\varrho = \xDD{\mu}{\mu^\circ}$ satisfies $0<c_{\varrho 0}\leq \varrho(x) \leq c_{\varrho 1}$ almost everywhere. 
\end{assumption}

On the smaller Sobolev--Hilbert space $\mc{G} = \mc{H}^s(\mbb{X}) \subset \mc{L}^2(\mbb{X})$ (for some $s>d_x/2$), there is no guarantee that $\mbb{Sp}(K)\subset \partial\mbb{D}$. 
It is possible that when the dynamics is chaotic, for some $g\in \mc H^s(\mbb{X})$, the Sobolev norm of $K^t g$ keeps growing due to the expansion in the magnitude of derivatives. 
We cannot even guarantee that a spectral resolution similar to \eqref{eq:spectral.resolution} exists, since $K$ may not be a \emph{normal} operator. 
Although one can verify that 
$$ \begin{aligned}
    \ip{\vk(x, \cdot)}{(KK^\ast - K^\ast K)\vk(x', \cdot)} =  \vk(f(x), f(y)) - \ip{\vk(x, f(\cdot))}{\vk(x', f(\cdot))}, 
\end{aligned}$$
whether the right-hand side is nonzero is not easy to test. In this case, we only know that $K$ and $K^{-1}$ are bounded linear operators due to Assumption \ref{assum:backward}.

\subsection{KKL observer construction}
Now we return to the problem of KKL observer. 
As we have reviewed in the Introduction, the theoretical problem of the existence of KKL observers have been extensively discussed, both for continuous-time systems and for discrete-time systems \cite{tran2023arbitrarily}. The numerical computation of the KKL observer, namely that of the injective mapping $\zeta$ and its left pseudoinverse, however, is nontrivial. 
The computation of the KKL injection, despite the existence result under mild backward distinguishability conditions, may involve intractable information constructed in the theoretical proof techniques.\footnote{For example, as seen in \cite{bernard2018luenberger}, the analytical expression of $\zeta$ requires the backward time since which the trajectory enters an enlarged domain from $\mbb{X}$, on which the dynamics is modified with a Friedrichs' mollifier. 
Only in the cases where the discrete-time dynamics $f$ is Lipschitz and invertible with a Lipschitz-like inverse \cite{brivadis2019luenberger} and $h$ is bounded on $\mbb{X}$, the desired injection $\zeta$ has an explicit series formula as shown in \eqref{eq:Brivadis.series}.
} 
In this work, since we aim at data-driven observer synthesis, it will be further assumed that $h$ is in the RKHS\footnote{This is by no means a restrictive assumption, since any $\mc{C}^s$ function is automatically in $\mc{H}^s$, and any continuously function can be approximated by a function with $\mc{C}^s$ regularity (e.g., by convolution with a smooth kernel).}, which implies that $h$ is bounded.
\begin{assumption}\label{assum:h}
    $h \in \mc{H}^s(\mbb{X})$. 
\end{assumption}

\par Second, the choice of parameters in the KKL observer can be restrictive. In \cite{brivadis2019luenberger} and \cite{tran2023arbitrarily}, the $A$ matrix in the observer must have a short enough time constant vis-{\`{a}}-vis the Lipschitz constant of $f$. When the model equations are not known precisely, the Lipschitz constant is difficult to estimate and tends to be overestimated. Actually, this Lipschitz constant-based tuning may not be necessary. 
In \cite{brivadis2023further}, for continuous-time systems with backward invariance, the KKL observer can be designed as a so-called ``deep KKL'' structure, meaning that the components of $\zeta = (\zeta_0, \dots, \zeta_{m-1})$ can be constructed as the outputs of a cascade of filters acting on the $y=h(x)$ signal. 
Here we follow a similar approach in the discrete-time setting. 
\begin{definition}\label{def:deep.KKL}
	A \emph{deep KKL observer} of order $m$ and parameter $\beta\in[0,1) \subset \mbb{D}$ is the following observer, with $k \in \mbb{Z}_{1,m-1}$:
	\begin{equation}\label{eq:deep.KKL}
		\begin{aligned}
			\zeta_0(\beta, f(x)) &= \beta \zeta_0(\beta, x) + (1-\beta) h(x) , \\
			\zeta_k(\beta, f(x)) &= \beta \zeta_k(\beta, x) + (1-\beta) \zeta_{k-1}(x).
		\end{aligned}
	\end{equation}
\end{definition}

\par Formally, we can resort to an expansion in backward time to derive an explicit formula for the first equation: $\zeta_0(\beta, x) = \sum_{\ell=0}^\infty \beta^\ell (1-\beta) h\bra{f^{-(\ell+1)}(x)}$. If the series converges, then we know that $\zeta_0: \mbb{D}\times \mbb{X} \rightarrow \mR$ is bounded and holomorphic with respect to $\beta$. 
By induction, for all $k\in \mbb{Z}_{0,m-1}$, 
$$  \zeta_k(\beta, x) = \sum_{\ell=0}^\infty \mr{C}_{k+\ell}^\ell \beta^\ell (1-\beta)^{k+1} \underbrace{h(f^{-(\ell+k+1)}(x))}_{K^{-(\ell+k+1)}h(x)} . $$
In fact, this arises from the Laurent expansion of the function $\lambda \mapsto ((1-\beta)/(\lambda-\beta))^{k+1}$ for $\lambda\in \partial\mbb{D}$:
$$\bra{ \frac{1-\beta}{\lambda-\beta}}^{k+1} = \sum_{\ell=0}^\infty \mr{C}_{k+\ell}^\ell \beta^\ell (1-\beta)^{k+1} \lambda^{-(k+\ell+1)}. $$
\begin{proposition}
    When $|\beta| < 1/\|K^{-1}\|_{\mc H^s\to \mc H^s}$, all $\zeta_k$ are bounded functions of $x$ on $\mc{H}^s(\mbb X)$ and holomorphic with respect to $\beta$. 
    When $|\beta| < 1$, all $\zeta_k\in \mc{L}^2(\mbb X)$. 
\end{proposition} 

\par Next, we prove that the deep KKL observer in Definition \ref{def:deep.KKL} is indeed a well-defined observer, with $\zeta = (\zeta_0, \dots, \zeta_{m-1})$ being \emph{injective} on $\mbb{X}$. 
\begin{proposition}
    The functions $\zeta_k$, $k\in \mbb{Z}_{0,m-1}$ satisfy the following recursive relation:
	$$ \xPP{}{\beta}\bra{ \frac{\zeta_k(\beta, x)}{k!(1-\beta)^{k+1}} } = \frac{\zeta_{k+1}(\beta, x)}{(k+1)!(1-\beta)^{k+2}}.$$
\end{proposition}
\begin{proof}
Indeed, by writing
$$\begin{aligned}
    \zeta_k(\beta, \cdot) =& (1-\beta)^{k+1} h\circ f^{-(k+1)} + \sum_{\ell=1}^\infty \mr{C}_{k+\ell}^\ell \beta^\ell (1-\beta)^{k+1} h\circ f^{-(k+\ell+1)} \\
    =& (1-\beta)^{k+1} h\circ f^{-(k+1)} 
    + \sum_{\ell=0}^\infty \mr{C}_{k+\ell+1}^{\ell+1} \beta^{\ell+1} (1-\beta)^{k+1} h\circ f^{-(k+\ell+2)} \\
\end{aligned}$$
and calculating 
$$\begin{aligned}
    \xPP{}{\beta} \bra{ \frac{\zeta_k(\beta, \cdot)}{(1-\beta)^{k+1}} } &= \sum_{\ell=0}^\infty \mr{C}_{k+\ell+1}^{\ell+1} (\ell+1) \beta^\ell h\circ f^{-(k+\ell+2)} \\
        & = \sum_{\ell=0}^\infty \mr{C}_{k+\ell+1}^{\ell} (k+1) \beta^\ell h\circ f^{-(k+\ell+2)} 
        = \frac{(k+1)\zeta_{k+1}(\beta, x)}{(1-\beta)^{k+2}}, 
\end{aligned}$$
the conclusion immediately follows. 
\end{proof}

\begin{assumption}\label{assum:distinguishability}
	System \eqref{eq:system} is backward distinguishable in a uniform time. That is, $\exists -m\in \mbb{Z}$ such that $-m<0$ and $\forall x, x'\in \mbb{X}$ with $x\neq x'$, $h\bra{f^{-m}(x)} \neq h\bra{f^{-m}(x')}$.
\end{assumption}
\begin{theorem}\label{thm:injectivity}
	Suppose that (i) $f$ invertible on $\mbb{X}$ and $f^{-1}$ is continuous, (ii) $h$ is bounded on $\mbb{X}$, and that (iii) Assumption \ref{assum:distinguishability} hold. Then for any fixed $\beta\in (0, 1)\subset \mbb{D}$, the mapping $\zeta(\beta, \cdot): \mbb{X}\rightarrow \mbb{R}^m$ is injective. 
\end{theorem}

\begin{proof}
	Denote $\tilde{\zeta}_k(\beta, x) = \tilde{\zeta}(\beta, x)/k!(1-\beta)^{k+1}$. Then by the proposition before, $\partial_\beta \tilde{\zeta}_k = \tilde{\zeta}_{k+1}$ holds for all nonnegative integers $k$. Define $\eta(\beta, x, x') = \zeta_0(\beta, x) - \zeta_0(\beta, x')$. As a function of $\beta\in \mbb{D}$, $\eta$ is holomorphic. We note that if $\partial_\beta^k \eta(\beta, x, x') = 0, \, \forall k\in \mbb{Z}_{0,m-1}$, then evaluation at $\beta=0$ implies that $x$ and $x'$ are not backward distinguishable within $m$ time steps. 
    Hence, define
    $$\mbb{K}_\ell = \BRA{ (x, x') \in \mbb{X}: \, x \neq x', \partial_\beta^k \eta(\beta, x, x') = 0 , \, \forall k\in \mbb{Z}_{0,\ell-1} }. $$
    Clearly, $\{\mbb{K}_\ell\}$ indexed by $\ell$ is a decreasing sequence of closed sets. By the assumption above, $\mbb{K}_m = \emptyset$. In other words, there does exist a distinct pair of states $x, x'\in \mbb{X}$ that satisfy $\partial_\beta^k \eta(\beta, x, x') = 0$ for all $k\in \mbb{Z}_{0, m-1}$, namely such that $\zeta_k(x)=\zeta_k(x')$ for all $k\in \mbb{Z}_{0, m-1}$ (according to the previous lemma). Therefore, the $\zeta$ mapping is injective.\footnote{
    The work of \cite{brivadis2023further} proved a similar theorem as our succeeding theorem for continuous-time systems, where the authors seem to assume a ``uniform'' backward time although the wording therein does not appear so. It is reasonable to postulate that such a uniformity is necessary. Because the compact subset $\mbb{K}$ is needed to establish the existence of an order $m$, the order $m$ must depend on the compact subset itself.} 
\end{proof}

With the above theorem that establishes the injectivity of $\zeta$, we reach at the expressions of KKL observers on RKHS. 
\begin{theorem}\label{cor:KKL.RKHS}
	Under Assumptions \ref{assum:backward}, \ref{assum:measure-preserving}, \ref{assum:h}, and \ref{assum:distinguishability} hold. Then the deep KKL observer \eqref{eq:deep.KKL} can be written as 
	\begin{equation}\label{eq:KKL.RKHS}
		\begin{aligned}
		z_{t+1} &= A_\beta z_t	+ b_\beta y_t, \\
		\hat{x}_t &= \zeta^\dagger(z_t).
		\end{aligned} 
	\end{equation}
	where $$A_\beta = \begin{bmatrix}
		\beta & 0 & \cdots & 0 & 0 \\
		1-\beta & \beta & \cdots & 0 & 0 \\
		\vdots & \vdots & \ddots & \vdots & \vdots \\
		0 & 0 & \cdots & 1-\beta & \beta 
	\end{bmatrix}, \enskip b_\beta = \begin{bmatrix}
	1-\beta \\ 0 \\ \vdots \\ 0
	\end{bmatrix}, $$ 
	The injection mapping $\zeta=(\zeta_0, \dots, \zeta_{m-1}) \in [\mc H^s(\mbb{X})]^m$ can be expressed in the Koopman operator form as
    \begin{equation}\label{eq:KKL.injection.operator.form}\zeta_k(x) = \sum_{\ell=0}^\infty \mr{C}_{k+\ell}^\ell \beta^\ell(1-\beta)^{k+1} \underbrace{ K^{-(k+\ell+1)}h(x) }_{ \ip{h}{(K^{\ast})^{-(k+\ell+1)}\vk(x, \cdot)} } 
	\end{equation}
    for $k\in \mbb{Z}_{0,m-1}$ and $x\in \mbb{X}$, where the underbraced formula is comprehended only in $\mc{N}_\kappa(\mbb{X})$ and the original formula can also be used in $\mc{L}^2(\mbb{X})$. 
\end{theorem}

\subsection{A non-chaotic example}
In this subsection, we illustrate the KKL observer with a simple example, where the system evolves on a limit cycle. 
The purpose is to show that if the Koopman spectrum on $\mc{L}^2(\mbb{X})$ is perfectly known, the observer indeed guarantees a vanishing observation error. Thus generally, the recovery of spectral information of $K|_{\mc{L}^2 \to \mc{L}^2}$ from data is helpful for observer synthesis. 
Indeed, by recalling the spectral resolution \eqref{eq:spectral.resolution} in the first formula of \eqref{eq:KKL.injection.operator.form}, we obtain 
\begin{equation}\label{eq:KKL.injection.spectral.form}
     \zeta_k = \sum_{\ell=0}^\infty \mr{C}_{k+\ell}^\ell \beta^\ell(1-\beta)^{k+1} \int_{\partial\mbb D} \lambda^{-(k+\ell+1)} \, \xD{E(\lambda)}h. 
\end{equation}

\par Consider a continuous-time system $\xD{x(t)}/\xD{t} = f_{\ms{c}}(x(t))$ on $\mR^2$ with a sampling time of $1$, where 
$$f_{\ms{c}}(x) = \begin{bmatrix} \alpha (1-x_1^2-x_2^2)x_1 - \gamma x_2 \\ \alpha (1-x_1^2-x_2^2)x_2 + \gamma x_1 \end{bmatrix}. $$
Here $\alpha > 0$ and $\gamma/2\pi\in(0, 1/4)$ is irrational. This system exhibits a limit cycle behavior. The invariant set of the system is the unit circle $\mbb{X} = \mbb{S}^1 = \{x\in \mR^2: \|x\| = 1\}$ on the real plane $\mR^2$, and the invariant measure $\mu$ can be taken as the arc length measure thereon. 
Clearly, the Koopman operator $K$ has a discrete spectrum, with eigenvalues $\lambda_p = e^{\mr{i}p\gamma}$ and associated eigenfunctions $\varphi_p(x) = e^{\mr{i}p\theta}$, where $\theta \in [-\pi, \pi)$ is the angle representation such that $x = (\cos\theta, \sin\theta)$, $p\in \mbb{Z}$. 
In this case, the spectral resolution can be expressed as $E(\{\lambda_p\}) = E_p$, $p\in \mbb{Z}$, where $E_p$ is the projection onto $\mr{span}\{\varphi_p\}$. 

\par
\begin{figure}[!t]
	\begin{center}
		\includegraphics[width=0.75\columnwidth]{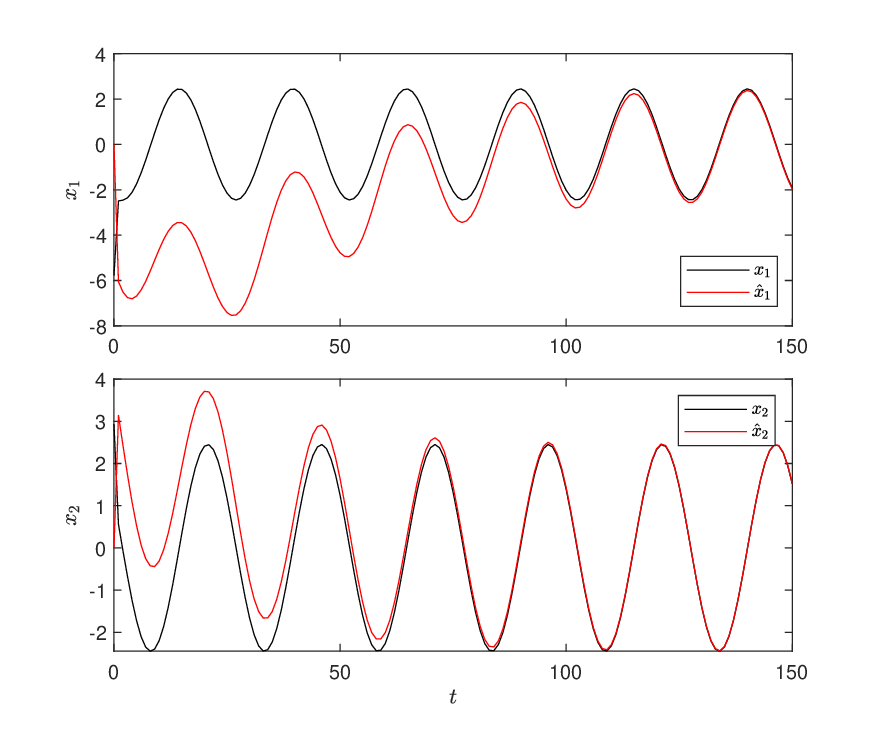} 
		\caption{KKL observer for a system with a rotation dynamics on the circle.} 
		\label{fig:rotation} 
	\end{center}
\end{figure}

 Therefore, if $h$ can be written as a Fourier series: $h = \sum_{p\in \mbb{Z}} c^h_p \varphi_p$, then the observer in \eqref{eq:KKL.injection.spectral.form} is reduced to:
$$ \zeta_k(\beta, x) = \sum_{p\in \mbb{Z}} \bra{\frac{1-\beta}{e^{\mr{i} p\gamma} -\beta}}^{k+1} c^h_p\varphi_p(x), \enskip k\in \mbb{Z}_{0,m-1}.$$
For a simple illustration, suppose that $h(x) = 2x_1$. Then $h(x) = \varphi_1(x) + \varphi_{-1}(x)$. Choosing $m=2$, we obtain
$$\begin{aligned}
	\zeta_0(\beta, x) &= \frac{1-\beta}{e^{\mr{i} \gamma} - \beta} e^{\mr{i}\theta} + \frac{1-\beta}{e^{-\mr{i} \gamma} - \beta} e^{-\mr{i}\theta},  \\
	\zeta_1(\beta, x) &= \bra{\frac{1-\beta}{e^{\mr{i} \gamma} - \beta}}^2 e^{\mr{i}\theta} + \bra{\frac{1-\beta}{e^{-\mr{i} \gamma} - \beta}}^2 e^{-\mr{i}\theta}.
\end{aligned}$$
The injectivity of the mapping from $\theta\in \mR\backslash 2\pi\mbb{Z}$ to $(\zeta_0(\beta, x), \zeta_1(\beta, x))$ is clear. Consider an example with $\alpha=1/5$ (radians), $\gamma=1/4$, and let $\beta=0.95$, we obtain an KKL observer. The performance of the KKL observer in a simulation is shown in Figure \ref{fig:rotation}. As $t\rightarrow \infty$, the state estimation error $\hat{x}-x$ approaches $0$. 

\par This example presented here is special, since the Koopman spectrum is discrete with explicit eigenfunctions and even the pseudo-inverse of the injective mapping can be found analytically. 
Generally, the Koopman spectrum may be continuous in chaotic cases, the spectral family $E$ may be difficult to characterize, and the pseudo-inverse does not have an explicit form. 
Next, we shall discuss how the observer can be computed numerically using data collected from the system.

\section{Data-Driven Computation of KKL Observer}\label{sec:computation}
\par Three different approaches are presented in this section for the approximate computation of the KKL observer, depending on the quality of available dataset. 
We always assume that $h$ is known. This is a reasonable assumption -- typically, the measurement mechanism is simple (e.g., the measured output is a state component) and clear to the user. However, $f$ is not known, and the knowledge of $f$ shall be compensated by the data collected on the state space. 
\begin{enumerate}
    \item In the first setting, a large number ($n$) of orbits are given. Each orbit ($i\in \mbb{Z}_{1,n}$) contains a point $x^{(i)}\in \mbb{X}$ and its past history up to time $-\ell$, namely $f^{-t}\bra{x^{(i)}} =: x^{(i, -t)}$, $t\in \mbb{Z}_{0,\ell}$. These sample points $x^{(i)}$ are independently drawn from $\mbb{X}$ under a probability measure $\nu$. 
    \item In the second setting, a single long orbit is given, which comes from the simulation of the system from an initial point $x^{(0)}$ to time $n-1$, that is, $x^{(t)} = f^t\bra{x^{(0)}}$ for $t\in \mbb{Z}_{0, n-1}$, resulting in $n$ data points. Such a long orbit is supposed to densely cover the attractor set $\mbb{X}$. 
    \item Lastly, we consider the ``snapshot'' setting, where a large number ($n$) of successor-predecessor pairs $\bra{x^{(i)}, f^{-1}\bra{x^{(i)}} }$ are given. These $x^{(i)}$ are independent chosen by a probability measure $\nu$. 
\end{enumerate}
The three settings provide different approaches for the estimation of the KKL injection $\zeta$. While the key goal of the observer computation is the estimation of the $\zeta^\dagger$ mapping, the procedure that we take to approximate $\zeta^\dagger$ from the estimated $\zeta$ is a convex routine -- kernel ridge regression.

\subsection{Estimation using a dataset of many orbits}
Given the data $x^{(i, -t)} = f^{-t}\bra{x^{(i)}}$ ($i\in \mbb{Z}_{1,n}$, $t\in \mbb{Z}_{0, \ell})$, in view of the expression of the $\zeta$ mapping in \eqref{eq:KKL.injection.operator.form}, we have that 
$$ \zeta_k(x^{(i)}) = \sum_{t=0}^\infty \mr{C}_{k+t}^k \beta^t (1-\beta)^{k+1} h\bra{x^{(i, -(k+t+1))}}$$
for $k\in \mbb{Z}_{0,m-1}$. Since the data available contains history only up to $-\ell$, we approximate $\zeta_k(x^{(i)})$ by a truncation on the data:
\begin{equation}\label{eq:zeta.sample.truncation}
	\tilde{z}_k^{(i)} = \sum_{t=0}^{\ell-m} \mr{C}_{k+t}^k \beta^t (1-\beta)^{k+1} h\bra{x^{(i, -(k+t+1))}}
\end{equation}
and let $\tilde{z}^{(i)} = \bra{ \tilde{z}_0^{(i)}, \dots, \tilde{z}_{m-1}^{(i)} }$. Indeed, when $t$ is summed from $0$ to $\ell-m$, the delayed time $(k+t+1)$ that appears on the $x$ superscript in \eqref{eq:zeta.sample.truncation} is at most $(m-1) + (\ell-m) + 1 =\ell$. 

\par The error resulted from this truncation turns out to be bounded, and such a bound decreases with increasing length of the orbits, $\ell$, at a nearly exponential rate, as shown in the following lemma. 
We may write the conclusion formally as $\varepsilon_\ell^{\ms{trunc}} \lesssim (\ell-1)^{m-1}\beta^\ell$, which converges to $0$ as $\ell\rightarrow \infty$. 

\begin{lemma}\label{lem:error.zeta.sample.truncation}
	Suppose that $0<\beta<1$ and $\tilde{\beta} = \beta\bra{1+\frac{m-1}{\ell-m}} < 1$, i.e., $\ell>m+\frac{m-1}{\beta^{-1} - 1}$. Then the truncation error satisfies:
    \begin{equation}\label{eq:error.zeta.sample.truncation}
		\norm{\tilde{z}^{(i)} - \zeta\bra{x^{(i)}}} \leq \mr{C}_{\ell-1}^{m-1} \sqrt{m}\upsilon_h \frac{1-\beta}{1-\tilde{\beta}} \tilde{\beta}\beta^{\ell-m}, 
	\end{equation}
	where $\upsilon_h = \sup \{|h(x)|: x\in \mbb{X}\}$. 
\end{lemma}

\begin{proof}
    By performing the truncation as in \eqref{eq:error.zeta.sample.truncation}, the error is bounded by: 
    \begin{equation*}
        \left\lvert \tilde{z}_k^{(i)} - \zeta_k\bra{x^{(i)}} \right\rvert \leq \sum_{t=\ell-m+1}^\infty \mr{C}_{k+t}^k \beta^t(1-\beta)^{k+1} \left\lvert h\bra{x^{(i, -(k+t+1))}} \right\rvert.
    \end{equation*}
    Bounding $|h(\cdot)|$ by $\upsilon_h$, $(1-\beta)^{k+1}$ by $1-\beta$ for all $k\in \mbb{Z}_{0,m-1}$, we have 
    \begin{equation*}
         \left\lvert \tilde{z}_k^{(i)} - \zeta_k\bra{x^{(i)}} \right\rvert \leq \upsilon_h(1-\beta) \sum_{t=\ell-m+1}^\infty \mr{C}_{k+t}^k \beta^t .
    \end{equation*}
    Since 
    \begin{equation*}
          \frac{\mr{C}_{k+t+1}^k}{\mr{C}_{k+t}^k} = \frac{k+t+1}{t+1} \leq 1+\frac{m-1}{\ell-m} = \frac{\tilde{\beta}}{\beta}, 
    \end{equation*}
    we know that 
    \begin{equation*}
      \left\lvert \tilde{z}_k^{(i)} - \zeta_k\bra{x^{(i)}} \right\rvert \leq \upsilon_h(1-\beta)\mr{C}_{k+\ell-m}^k \sum_{t=\ell-m+1}^\infty (\tilde{\beta}/\beta)^{t-\ell+m} \beta^t .
    \end{equation*}
    Hence, by upper-bounding over $k\in \mbb{Z}_{0,m-1}$, we have 
    \begin{equation*}
     \left\lvert \tilde{z}_k^{(i)} - \zeta_k\bra{x^{(i)}} \right\rvert \leq \upsilon_h (1-\beta)\mr{C}_{\ell-1}^{m-1} \frac{\tilde{\beta}\beta^{\ell-m}}{1-\tilde{\beta}} .
    \end{equation*} 
    This bound is on all component indices $k$. The conclusion then follows. 
\end{proof}

\par Then, we can use the collection of pairs $\BRA{x^{(i)}, \tilde{z}^{(i)}}_{i=1}^n$ as the sample data, each satisfying the approximate relation $\tilde{z}^{(i)}\approx \zeta(x^{(i)})$ in the sense of \eqref{eq:error.zeta.sample.truncation}, to estimate $\zeta^\dagger$. 
As a nonlinear mapping in general, a useful approach for this estimation is \emph{kernel ridge regression} (KRR), where we define a Mercer kernel $\kappa$ on $\mR^m$ (as $m$ is the dimension of $z$) and seek a function of the following form in the RKHS $\mc{N}_\kappa$: 
\begin{equation}\label{eq:KRR.function.form}
	 \widehat{\zeta_k^\dagger} = \sum_{i=1}^n c_{ki}\kappa\bra{\tilde{z}^{(i)}, \cdot}, \enskip k\in \mbb{Z}_{1,d_x}. 
\end{equation}
To determine the coefficients $c_{ki}$ (separately over state components $k\in \mbb{Z}_{1,d_x}$), we solve the following optimization problem that minimizes the mean-squared error, while penalizing the RKHS norm with a regularization parameter $\alpha$:
\begin{equation*}
    \min_{c_{ki}} \sum_{j=1}^n \bra{\sum_{i=1}^n c_{ki}\kappa\bra{\tilde{z}^{(i)}, \tilde{z}^{(j)}} - x_k^{(j)} }^2 + \frac{\alpha}{2}\norm{\sum_{i=1}^n c_{ki} \kappa\bra{\tilde{z}^{(i)}, \cdot} }_{\mc{N}_\kappa}^2. 
\end{equation*}

\noindent Denote $c_k = (c_{k1}, \dots, c_{kn})$, $x_k = (x_k^{(1)}, \dots, x_k^{(n)})$, and the kernel matrix $G_\kappa \in \mR^{n\times n}$ whose $(i,j)$-th entry is $\kappa\bra{\tilde{z}^{(i)}, \tilde{z}^{(j)}}$. The minimizer has an explicit form:
\begin{equation}\label{eq:KRR.optimization}
	\hat{c}_k = \bra{G_\kappa c_k +\alpha I}^{-1} x_k. 
\end{equation}

\par To analyze the error of KRR, essentially we consider the problem as that of recovering a vector-valued function $\zeta^\dagger$ with input ``$z$'' and output ``$x$'', where the \emph{input} (instead of the output) is collected with error. Such a problem differs from the usual setting in the theory of KRR, where a description of the error in the output data is needed to establish the KRR error bound \cite{smale2007learning}. 
To resolve this issue, we have to rely on some further conditions on $\zeta$ and $\zeta^\dagger$. 

\begin{theorem}\label{thm:guarantee.1}
	Let $\varepsilon>0$. Suppose that 
	\begin{enumerate}[label=(\roman*)]
		\item the conditions needed for Theorem \ref{thm:injectivity} all hold true;
		\item $\zeta^\dagger$ can be extended to $\zeta(\mbb{X})+\bar{\mbb{B}}_\varepsilon = \{\zeta(x)+\varepsilon_z: x\in \mbb{X}, \varepsilon_z\in \mR^m, \|\varepsilon_z\|\leq \varepsilon\}$ as a Lipschitz continuous function with Lipschitz constant $c_{\zeta^\dagger}$; 
		\item $\ell$ is sufficiently large so that $\tilde{\beta}= \beta \bra{1+\frac{m-1}{\ell-m}} < 1$ and the right-hand side of \eqref{eq:error.zeta.sample.truncation} does not exceed $\varepsilon$; and, 
		\item $\zeta^\dagger$, component-wise, lies in the range of $L_\kappa: \mc{L}^2_{\nu_z} \rightarrow \mc{N}_\kappa$, where we denote $\nu_z = \zeta_\sharp \nu$, namely the corresponding distribution of $\zeta(x)$ when $x$ is sampled from the given $\nu$, and $L_\kappa$ is the integral operator defined by $L_\kappa \phi(z) = \int_{\zeta(\mbb{X})} \kappa(z,z') \phi(z') \xD{\nu_z(z)}$. 
	\end{enumerate} 
	Then, with a proper choice of $\alpha$, there exists a constant $c_{\ms{KRR}}>0$ such that with a confidence of $1-\delta$ over the draw of sample points, the following bound holds uniformly on $\mbb{X}$: 
	\begin{equation}\label{eq:error.zeta.sample.KRR}
		\norm{ \widehat{\zeta^\dagger} (\zeta(x)) - x } \leq \frac{c_{\ms{KRR}}}{n^{1/6}} \log \frac{2}{\delta} + c_{\zeta^\dagger}\varepsilon 
	\end{equation}
	for the $\widehat{\zeta^\dagger}$ returned from Algorithm \ref{alg:1}. 
\end{theorem}

\begin{proof}
    Using condition (i), $\zeta^\dagger$ exists on $\zeta(\mbb{X})$. Under condition (iii), $\tilde{\zeta}$ deviates from $\zeta$ by a distance of at most $\varepsilon$, and hence by using condition (ii), $\widehat{\zeta^\dagger}$ is a (suboptimal) solution to the KRR problem. 
    According to \cite{smale2007learning} (Corollary 2 therein), under condition (iv), with confidence $1-\delta$, we have 
    $$  \norm{\widehat{\zeta^\dagger} - \zeta^\dagger} \leq 4\log\frac{2}{\delta} \bra{3\sup_{x\in \mbb{X}} \|x\|}^{1/3} \norm{ L_\kappa^{-1}\zeta^\dagger}_{\mc{L}^2_{\nu_z}}^{2/3} \frac{1}{n^{1/6}}. $$
    The norm on $\widehat{\zeta^\dagger} - \zeta^\dagger$ is a sup-norm on $\tilde{\zeta}(\mbb{X})$, which is contained in $\zeta(\mbb{X}) + \bar{\mbb{B}}_\varepsilon$. The right-hand side can be denoted by $c_{\ms{KRR}}n^{-1/6} \log(2/\delta)$. Hence, for all $z\in \tilde{\zeta}(\mbb{X})$, 
    $$ \norm{\widehat{\zeta^\dagger} (z) - \zeta^\dagger(z) } \leq \frac{c_{\ms{KRR}}}{n^{1/6}} \log\frac{2}{\delta}.$$
    For all $x\in \mbb{X}$, there exists a $z\in \tilde{\zeta}(\mbb{X})$ that differs from $\zeta(x)$ by a distance at most $\varepsilon$. The conclusion thus follows.
\end{proof}

\par Up to this point, \eqref{eq:error.zeta.sample.KRR} gives an upper bound on the state observation error, which vanishes as the number of backtracked time steps $\ell$ tends to infinity and then the sample size $n$ tends to infinity. 
\RestyleAlgo{ruled}
\begin{algorithm}[!t]
	\caption{Data-driven synthesis of KKL observer given many orbits}\label{alg:1}
	\KwData{$\BRA{f^{-t}\bra{x^{(i)}} =: x^{(i, -t)}: t\in \mbb{Z}_{-\ell,0}, i\in \mbb{Z}_{1,n}}$}
	\KwResult{Mapping $\widehat{\zeta^\dagger}$}
	Set observer order $m$ and parameter $\beta\in (0,1)$\; 
	Set kernel $\kappa$ and $\alpha>0$ for KRR\;
	\For{$i\in \mbb{Z}_{1,n}$}{
		Calculate $\tilde{z}^{(i)}$ according to \eqref{eq:zeta.sample.truncation}\;
	}
	Calculate kernel matrix $G_\kappa$\;
	\For{$k\in \mbb{Z}_{1,d_x}$}{
		Calculate $c_k$ according to \eqref{eq:KRR.optimization}\;
		Let $\widehat{\zeta_k^\dagger}$ be given as in \eqref{eq:KRR.function.form}\;
	}  
\end{algorithm}

\begin{remark}
    Among the conditions in Theorem \ref{thm:guarantee.1}, condition (ii) actually requires additional ``strictness'' of injection $\zeta$; specifically, if $\|\zeta(x)-\zeta(x')\|\geq c\|x-x'\|$ for any $x,x'\in \mbb{X}$ (i.e., if $\zeta$ is strictly injective), then its left pseudo-inverse $\zeta^\dagger$ satisfies $\|\zeta^\dagger(z)-\zeta^\dagger(z')\|\leq c^{-1}\|z-z'\|$ for any $z,z'\in \zeta(\mbb X)$ (i.e., is Lipschitz). 
    The extension from $\zeta(\mbb X)$ to $\zeta(\mbb X) + \bar{\mbb{B}}_\varepsilon$ requires that $\zeta(\mbb X)$ is an open set, which can result from the continuity of $\zeta$.  
\end{remark}

\begin{remark}
    The conclusion in Theorem \ref{thm:guarantee.1} is one regarding the uniform error bound on $\widehat{\zeta^\dagger}$. Apart from the truncation error $\varepsilon$ that depends on the truncation length $\ell$, the generalized error scales down with increasing sample size at a rate of $n^{-1/6}$. 
    This rate is associated with the extent of regularity assumed on $\zeta^\dagger$; that is, $\zeta^\dagger$ is in the range of $L_\kappa$ componentwise. 
    If $\zeta^\dagger$ is assumed to lie in the range of $L_\kappa^r$ for some $r\in (1/2, 1]$, then the rate further deteriorates to $n^{-(2r-1)/(4r+2)}$ \cite{smale2007learning}. 
    \par If we bound the generalized error in $\mc{L}^2$ (i.e., mean-squared error) instead of uniformly, then the rate can be much improved. 
    Using Theorems 2.3 and 3.1 of \cite{cucker2007learning}, we reach at the conclusion that if $\kappa$ is a Gaussian kernel, then there exist constants $c_0, c_1, c_2$ such that 
    $$\norm{\widehat{\zeta^\dagger} (\zeta(x)) - x }_{\mc{L}_{\nu_z}^2} \leq \varepsilon_{\ms{KRR}}+ c_{\zeta^\dagger}\varepsilon, $$
    in which $\varepsilon_{\ms{KRR}}$ varies with the sample size $n$ by
    $$n = \frac{1}{\varepsilon_{\ms{KRR}}^2}\Bra{c_0\bra{\log\frac{1}{\varepsilon_{\ms{KRR}}}}^{\frac{m}{2}} + c_1\log\frac{1}{\delta} + c_2}. $$
    Then, roughly speaking, the $\mc{L}^2$-error scales at a typical rate of $n^{-1/2}$. When the kernel $\kappa$ is of class $\mc{C}^s$ instead of being analytical, we have 
    $$n = \frac{1}{\varepsilon_{\ms{KRR}}^2}\Bra{\frac{c_0}{\varepsilon_{\ms{KRR}}^{2m/s}} + c_1\log\frac{1}{\delta} + c_2}, $$
    which is better than the scaling of the uniform bound whenever $s>m/2$. 
\end{remark}

\subsection{Estimation using a single long orbit}
If the user is not able to collect a (very dense) set of points and backtrack each of them for a very long time $\ell$, but instead is only able to simulate the system in forward time, we should consider the setting where the dataset is the orbit of length $n$ starting from a single point $x^{(0)}\in \mbb{X}$. 
When $n$ is large enough, we may envision that the points on the orbit $x^{(t)} = f^t\bra{x^{(0)}}$, $t\in \mbb{Z}_{0, n-1}$, cover the entire invariant set $\mbb{X}$ \emph{densely}. As such, we may approximate the function $h$ by the kernel function at these data points starting from a count of $\ell$: 
\begin{equation}\label{eq:KRR.h}
	 \hat{h}(\cdot) = \sum_{i=\ell}^{n-1} c^h_i\vk \bra{x^{(i)} , \cdot}. 
\end{equation}
The determination of the coefficients $c^h_i$  follows a KRR procedure with a low regularization parameter. While now the sample points are not independently sampled from a distribution, their density enables the \emph{fill distance}:
$$ \delta_{\ms{fill}} = \sup_{x\in \mbb{X}}\min_{i\in \mbb{Z}_{\ell, n-1}} \norm{ x-x^{(i)} } $$ 
to play a role in establishing the approximation error. 

\begin{lemma}[\cite{wendland2004scattered}, Theorem 11.17]
\label{lem:kernel.approximant}
	Suppose that $\mbb{X}$ satisfy an interior cone condition.\footnote{That is, there is an angle parameter $\theta\in (0,\pi/2)$ and a radius parameter $r$, such that for any point $x$ in the interior of $\mbb{X}$, there is a cone with vertex $x$, angle $\theta$, and radius $r$ contained in $\mbb{X}$.} 
    Then $\exists c_{\ms{fill}}>0$ such that
	$$|h(x)-\hat{h}(x)| \leq c_{\ms{fill}} \|h\|_{\mc{N}_\vk}  \delta_{\ms{fill}}^q, \enspace \forall x\in \mbb{X}, $$
	where $\mc{N}_{\vk}(\mbb{X}) \equiv \mc{H}^s(\mbb{X})$ with $s=q+d_x$. 
\end{lemma}

\par Now we use the action of $(K^\ast)^{-(k+\ell+1)} \vk(x,\cdot)$ on each $x^{(j)}$ to define a corresponding approximation of $\zeta$, denoted as $\tilde{\zeta} = (\tilde{\zeta}_0, \dots, \tilde{\zeta}_{m-1})$, by a truncation of the infinite series in \eqref{eq:KKL.injection.operator.form} in the same way as in the previous subsection (taking $t=0, \dots,\ell-m$) and substitution of $h$ by $\hat{h}$:
\begin{equation}\label{eq:zeta.sample.truncation.2}
    \tilde{\zeta}_k\bra{x^{(j)}} = \sum_{t=0}^{\ell-m} \mr{C}_{k+t}^t \beta^t(1-\beta)^{k+1} \underbrace{ \ip{ \hat{h} }{  (K^\ast)^{-(k+t+1)} \vk\bra{x^{(j)}, \cdot}} }_{ \sum_{i=\ell}^{n-1} c^h_i \vk\bra{x^{(i)}, x^{(j-k-t-1)}} }.
\end{equation}

\noindent It then follows that the approximation of $\zeta$ by $\tilde{\zeta}$ contains a truncation error and a fill distance-based error. That is, 
$\|\tilde{\zeta} - \zeta\| \leq \varepsilon^{\ms{trunc}} + \varepsilon^{\ms{fill}}$, 
where $\varepsilon^{\ms{trunc}}$ is as shown in \eqref{eq:error.zeta.sample.truncation} in the previous subsection and $\varepsilon^{\ms{fill}}$ results from the approximation of $h$ by $\hat{h}$ in \eqref{eq:zeta.sample.truncation.2}. Then
\begin{equation}\label{eq:error.zeta.sample.fill}
    \begin{aligned}
        \varepsilon^{\ms{fill}} &\leq \bra{ \bra{1+c_{\ms{fill}}} \delta^{q}_{\ms{fill}} + c_{\ms{fill}} \delta^{2q}_{\ms{fill}} } \sqrt{m} \max_{k\in \mbb{Z}_{0,m-1}} \sum_{t=0}^{\ell-m} \mr{C}_{k+t}^k \beta^t (1-\beta)^{k+1} \\
        & = c_{\ms{fill}}' \bra{\delta^{q}_{\ms{fill}} + \delta^{2q}_{\ms{fill}}}. 
    \end{aligned}
\end{equation}

\par Finally, with the approximated $\zeta$ mapping, the left pseudo-inverse $\widehat{\zeta^\dagger}$ is estimated via the same KRR procedure as in the previous subsection, where a dataset on the $x$-space according to distribution $\nu_x$ is drawn and $\tilde{\zeta}$ is evaluated on these data points. 
A similar theorem for the performance guarantee of KRR based on this new $\tilde{\zeta}$ can be obtained. 
\begin{theorem}\label{thm:guarantee.2}
	Let $\varepsilon>0$. Suppose that 
	\begin{enumerate}[label=(\roman*)]
		\item the conditions needed for Theorem \ref{thm:injectivity} all hold true;
		\item $\zeta^\dagger$ can be extended to $\zeta(\mbb{X})+\bar{\mbb{B}}_\varepsilon = \{\zeta(x)+\delta: x\in \mbb{X}, \|\delta\|\leq \varepsilon\}$ as a Lipschitz continuous function with Lipschitz constant $c_{\zeta^\dagger}$; and, 
		\item $\ell$ is sufficiently large so that $\tilde{\beta}= \beta \bra{1+\frac{m-1}{\ell-m}} < 1$, and the truncation error (i.e., the right-hand side of \eqref{eq:error.zeta.sample.truncation}) added to the fill distance-based error in \eqref{eq:error.zeta.sample.fill}, with a sufficiently large $n$, does not exceed $\varepsilon$; and furthermore, 
	\item $\zeta^\dagger$, component-wise, lies in the range of $L_\kappa: \mc{L}^2_{\nu_z} \rightarrow \mc{N}_\kappa$. 
	\end{enumerate} 
	Then, there exists $c_{\ms{KRR}}>0$ such that with a confidence of $1-\delta$ over the draw of sample points in KRR, the bound \eqref{eq:error.zeta.sample.KRR} holds uniformly on $\mbb{X}$ (with $n$ replaced by $n-\ell$) for the $\widehat{\zeta^\dagger}$ mapping obtained from Algorithm \ref{alg:2}. 
\end{theorem} 

\begin{algorithm}[!t]
	\caption{Data-driven synthesis of KKL observer given a single long orbit}\label{alg:2}
	\KwData{$\BRA{f^i\bra{x^{(0)}} =: x^{(i)}: i\in \mbb{Z}_{0,n-1}}$}
	\KwResult{Mapping $\widehat{\zeta^\dagger}$}
	Set observer order $m$ and parameter $\beta\in (0,1)$\; 
	Set truncation length $\ell$, $z$-kernel $\kappa$ and $x$-kernel $\varkappa$\; 
	Solve \eqref{eq:KRR.h} to determine $c^h_i$ in $\hat{h}$, $i\in\mbb{Z}_{\ell,n-1}$\;
	\For{$j\in \mbb{Z}_{\ell,n-1}$}{
		Calculate $\tilde{z}^{(j)} = \tilde{\zeta}\bra{x^{(j)}}$ by \eqref{eq:zeta.sample.truncation.2}\;
	}
	\For{$k\in \mbb{Z}_{1,d_x}$}{
		Calculate $c_k$ by \eqref{eq:KRR.optimization} with sample $\BRA{x^{(j)}, \tilde{z}^{(j)}}_{j=\ell}^{n-1}$\;
		Let $\widehat{\zeta_k^\dagger}$ be given as in \eqref{eq:KRR.function.form}\;
	}
\end{algorithm}

\subsection{Estimation using snapshots}
When only snapshots are provided in the dataset, it becomes infeasible to backtrack or simulate forward  on the data points. The structure of Koopman spectrum then should be found useful. Our intuitive ideas are as follows.
\begin{enumerate}
	\item If there is a linear combination of kernel functions $\{\vk(x^{(i)}, \cdot)\}_{i=1}^n$ that is an (approximate) eigenfunction of $K^\ast$ associated with eigenvalue $\lambda\in \partial\mbb{D}$, then the action of the inverse powers of $K^\ast$ on this eigenfunction will have explicit expressions that are still linear combinations of the kernel functions. 
	\item A spectrum point should be an ``approximate'' eigenvalue in a proper sense, associated with some ``approximate'' eigenfunctions. If the dataset is sufficiently large on $\mbb{X}$, these approximate eigenvalues should be rich enough to cover the Koopman spectrum and the corresponding approximate eigenfunctions should cover the RKHS. 
	\item Hence, at any sample point $x\in \mbb{X}$, by decomposing $\vk(x, \cdot)$ as a combination of such approximate eigenfunctions, an estimate of $\zeta$ can be obtained. The error bound should depend on the suitability of such ``approximate'' eigenvalues and eigenfunctions. 
\end{enumerate} 
To this end, we rewrite $(K^\ast)^{-1}$ (as an operator on $\mc L^2(\mbb X)$) in its spectral resolution and denote by $E'$ its spectral family:
$$ (K^\ast)^{-1} = \int_{\partial\mD} \lambda \xD{E'(\lambda)}. $$
As this operator may not even possess an eigenvalue at any spectrum point $\lambda \in \partial\mD$, we quantify the suitability of taking any given nonzero function $g\in \mc L^2(\mbb X)$ as an approximate eigenfunction by the \emph{residual}:
\begin{equation*}
    \ms{res}(g, \lambda) = \norm{((K^*)^{-1}-\lambda)g}_{\mc L^2}  \big{/} \|g\|_{\mc L^2}. 
\end{equation*}

\begin{lemma}\label{lem:residual}
    Suppose that $\mbb{J}\subset \partial\mD$ contains $\lambda$ and $E'(\mbb{J})\neq 0$. Then there exists a nonzero $\psi\in \mc{H}^s(\mbb{X}) \subset \mc L^2(\mbb{X})$ such that $\ms{res}(\psi, \lambda) \leq \delta = \inf_{\lambda'\in \mbb{J}}|\lambda' - \lambda|$. 
\end{lemma}

\begin{proof}
    Since $E'(\mbb J)\neq 0$, there exists $g\in \mc{L}^2(\mbb X)$ such that $E'(\mbb J)g\neq 0$. For the spectral family $E'$, any two disjoint subsets of the spectrum, $\mbb{A}_1$ and $\mbb{A}_2$, satisfies $E'(\mbb{A}_1) E'(\mbb{A}_2) = 0$, and due to this, for $\psi = E'(\mbb J)g$, it holds that $\psi=E'(\mbb J)\psi$ and 
    $$ (K^\ast)^{-1} \psi = \int_{\partial\mD} \lambda \xD{E'(\lambda)} E'(\mbb{J})\psi = \int_{\mbb{J}} \lambda \xD{E'(\lambda)} \psi. $$
    Therefore, we have
    $$\begin{aligned}
    	\|((K^\ast)^{-1} - \lambda) \psi\|_{\mc{L}_2}^2 &= \norm{\int_{\mbb{J}} \bra{\lambda'-\lambda} \xD{E'(\lambda')} \psi}_{\mc{L}_2}^2 \\
    	&\leq \int_{\mbb{J}} |\lambda'-\lambda|^2 \xD{|\ip{\psi}{E'(\lambda') \psi}|}
    	\leq \delta^2 \|\psi\|_{\mc{L}_2}^2,  
    \end{aligned}$$
    i.e., $\ms{res}(\psi, \lambda) \leq \delta$. In the above construction, because $E'(\mbb J)$ is a projection, as long as $g\in \mc{H}^s(\mbb{X}) \subset \mc{L}^2(\mbb{X})$, $\psi = E'(\mbb J) g$ remains in $\mc{H}^s(\mbb{X})$. 
\end{proof}

Hence, we divide the unit circle $\partial\mD$ into $p$ disjoint intervals $\{\mbb J_j\}_{j=1}^p$ such that in each interval $\mbb{J}_j$, every point $\lambda$ has a distance from its center $\lambda_j$ bounded by $\delta_{\ms{mesh}}$. 
If the number of grid points $p$ are large enough, there should exist a function $\psi_j \in \mc{L}^2(\mbb{X})$ with a residual $\ms{res}(\psi_j, \lambda_j)\leq \delta_{\ms{mesh}}$; moreover, these $\psi_j$'s are mutually orthogonal in $\mc{L}^2(\mbb{X})$. 
To computationally approach every $\psi_j$, it is only possible to approximate it on the $n$-dimensional function space spanned by the kernel functions at the data points: $\mc{G}_n = \mr{span}\{\vk(x^{(i)}, \cdot)\}_{i=1}^n$. We thus need to rely on a large number of data points $n$ to achieve good approximation. 
That is, given $\{x^{(i)}\}_{i=1}^n$, we solve the following residual minimization problem for each $j=1,\cdots,p$:
\begin{equation}
    \label{eq:residual.minimization.L2}
    \min_{\hat \psi_j\in \mc G_n, \|\hat \psi_j\|_{\mc L^2}^2=1} \norm{((K^\ast)^{-1}-\lambda_j) \hat\psi_j}_{\mc L^2}^2. 
\end{equation}
Clearly, for each $\psi_j$, its kernel-based approximant in $\mc{G}_n$, denoted as $\hat\psi_j$ due to Lemma \ref{lem:kernel.approximant}, should satisfy $\|\psi_j - \hat\psi_j\|_{\mc L^2} \leq c_{\ms{fill}} \|h\|_{\mc N_\vk} \delta_{\ms{fill}}^q$, and hence, 
$$|\ms{res}(\hat\psi_j, \lambda_j) - \ms{res}(\psi_j, \lambda_j)| \leq \mr{const} \cdot \bra{\delta_{\ms{fill}}^q + \delta_{\ms{fill}}^{2q}}. $$
When $n$ is large so that $\delta_{\ms{fill}}$ is small, the second term is negligible compared to the first one, and hence $\ms{res}(\hat\psi_j, \lambda_j) \leq \delta_{\ms{mesh}} + c\delta_{\ms{fill}}^q$ with $c=\mr{const}$. 
Now that $\hat{\psi}_j$ are the minimizers of the residuals, they should have good $\mc L^2$-orthogonality. 
\begin{lemma}\label{lem:well-spanning}
    Suppose that $\lambda_1,\dots,\lambda_p$ are chosen as equally spaced points on $\partial\mbb{D}$. When $p$ is large enough, $\mr{span}\{\hat{\psi}_1, \dots, \hat{\psi}_p\} \supset \mc{G}_n$. 
\end{lemma}
\begin{proof}
	Let all $\hat{\psi}_j$ be normalized. We consider the inner products of $\hat{\psi}_j$ and $\hat{\psi}_{j'}$. Denote $\omega_j = ((K^{-1})^\ast-\lambda_j) \hat\psi_j$. Then $\|\omega_j\|\leq \delta_{\ms{mesh}} + c\delta_{\ms{fill}}^q=:\delta$. Since $\mbb{Sp}((K^{-1})^\ast) \subset \partial\mbb{D}$, 
	$$\ip{\hat{\psi}_j}{\hat{\psi}_{j'}} = \ip{(K^{-1})^\ast\hat{\psi}_j}{(K^{-1})^\ast\hat{\psi}_{j'}} = \ip{\lambda_j\hat{\psi}_j + \omega_j}{\lambda_{j'}\hat{\psi}_{j'} + \omega_{j'}}.$$
	Hence $$(1-\lambda_j\lambda_{j'}^\ast)|\ip{\hat{\psi}_j}{\hat{\psi}_{j'}}|\leq 2\delta+\delta^2.$$ 
	Without loss of generality, consider $p=2p'$ equally-spaced points on $\partial\mbb{D}$: $\lambda_j = \exp(\mr{i}\theta_j)$, where $\theta_j = \pi j/p'$, $j\in \mbb{Z}_{-(p'-1), p'}$. Then $1-\lambda_j \lambda_{j'}^\ast = \sin (\pi|j-j'|/2p')$ and hence 
	$$|\ip{\hat{\psi}_j}{\hat{\psi}_{j'}}|\leq \delta(1+\delta/2) \csc (\pi|j-j'|/2p'). $$
	Let the $2p'\times 2p'$ matrix formed by the inner products of the approximate eigenfunctions:
    $$\Psi = \begin{bmatrix}
        1 & \ip{\hat{\psi}_1}{\hat{\psi}_2} & \cdots & \ip{\hat{\psi}_1}{\hat{\psi}_{2p'}} \\
        \ip{\hat{\psi}_2}{\hat{\psi}_1} & 1 & \cdots & \ip{\hat{\psi}_2}{\hat{\psi}_{2p'}} \\
        \vdots & \vdots & \ddots & \vdots \\
        \ip{\hat{\psi}_{2p'}}{\hat{\psi}_1} & \ip{\hat{\psi}_{2p'}}{\hat{\psi}_2} & \cdots & 1
    \end{bmatrix}. $$
    Let $V = [v_1, \dots, v_{2p'}]$, each $v_j\in \mC^n$ defining the coefficients of $\hat{\psi}_j$ in $\{\vk(x_i,\cdot)\}_{i=1}^n$. Then $\Psi = V^\ast G_\vk V$, which is clearly a unitary matrix. 
    \par Suppose that $\mc{G}_n$ has a dimension of $r$. Let $2p'=rq'$ and we take a submatrix of $\Psi$, denoted as $\Psi_r$, by extracting every other $q'$ rows and columns. Then $\Psi$ has at least a rank of $r$ if $\Psi_r$ has a rank of $r$. By Gershgorin's circle theorem, $\Psi_r$ have full rank ($r$) if for every row, the off-diagonal elements have a sum of moduli lower than $1$. By elementary calculation of this sum, we obtain the following sufficient condition, where we for simplicity assume that $r=2r'$:
	$$ \frac{1}{\delta(1+\delta/2)} > 1 + 2\sum_{j=1}^{r'-1} \csc \frac{\pi j}{2r'},$$
	for which it suffices to have 
	$$ \frac{1}{\delta(1+\delta/2)} > 1 + \frac{4r'}{\pi}\int_{\pi/2r'}^{\pi/2}\csc x \xD{x} = 1 + \frac{4r'}{\pi}\log\cot \frac{\pi}{2r'}.$$
	There must be a sufficiently large $p'$ that satisfies the inequality, by adopting a large $q'$ with a fixed $r'$. 
\end{proof}

\begin{corollary}
	Let $g\in \mc{G} = \mc{N}_\vk(\mbb{X}) \equiv \mc{H}^s(\mbb{X})$ with $\|g\|_{\mc{N}_\vk} = 1$ (including the case of $g=\vk(x, \cdot)$ for some fixed $x\in \mbb{X}$), and denote its approximant on the basis of $\{\hat{\psi}_j\}_{j=1}^p$ as $\tilde{g}= \sum_{j=1}^p \tilde{c}^g_j \hat{\psi}_j$. Then if $p$ is sufficiently large, we have 
	$$\|g - \tilde{g}\|_{\mc{N}_\vk} \leq c_{\ms{fill}}\delta_{\ms{fill}}^{q}, \text{ and }
    \|g - \tilde{g}\|_{\mc{L}^2} \leq c_{\ms{fill}}'\delta_{\ms{fill}}^{q}$$
	where $q=s-d_x/2>0$ and $c_{\ms{fill}}, c_{\ms{fill}}'>0$ are constants. 
\end{corollary}

Therefore, by taking the KRR approximant of $\tilde{h}$ of $h$ in $\mc{G}_n$, which induces an error in $\mc{L}^2(\mbb{X})$ proportional to $\delta_{\ms{fill}}^q$ as established in the above corollary, we have the following approximate construction of $\tilde{\zeta}_k$, $k\in \mbb{Z}_{0,m-1}$:
\begin{equation}\label{eq:zeta.sample.truncation.3}
	\tilde{\zeta}_k\bra{x} = \sum_{t=0}^{\ell-m} \mr{C}_{k+t}^t \beta^t(1-\beta)^{k+1} \underbrace{\ip{ \tilde{h} }{ (K^\ast)^{-(k+t+1)} \vk(x, \cdot)} }_{ \sum_{j=1}^{p} \tilde{c}^h_j\lambda_j^{k+t+1} \ip{\hat{\psi}_j(\cdot)}{\vk(x, \cdot)}}.
\end{equation}
The propagation of the $\mc{L}^2$-errors under the summation over $t\in \mbb{Z}_{0, \ell-m}$ can be bounded in an analogous fashion to the previous subsections. As a result, the estimated $\tilde{\zeta}$ function has a bounded $\mc{L}^2$-error from the true $\zeta$. For each $x^{(i)}$ sampled from distribution $\nu$ on $\mbb{X}$, the deviation of $\tilde{\zeta}(x^{(i)})$ from $z^{(i)} = \zeta(x^{(i)})$ has a bounded variance. 
Now that $\zeta^\dagger$ is to be found by regression on $(\tilde{z}^{(i)},x^{(i)})$-data, the Lipschitz assumption of $\zeta^\dagger$ guarantees that $\|x^{(i)} - \zeta^\dagger(z^{(i)})\|\leq c_{\zeta^\dagger} \|z^{(i)} - \tilde z^{(i)}\|$ bounded in $\mc{L}^2$ as well, and in addition $x^{(i)}\in \mbb{X}$ -- a bounded region. The conclusion of \cite{smale2007learning} remains instrumental. 
Finally, we arrive at the following conclusion. 

\begin{theorem}\label{thm:guarantee.3}
	Let $\varepsilon>0$. Suppose that the conditions of Theorem \ref{thm:guarantee.2} hold. Then the conclusions of Theorem \ref{thm:guarantee.2} hold for the $\widehat{\zeta^\dagger}$ mapping obtained from Algorithm \ref{alg:3}. 
\end{theorem}

\begin{algorithm}[!t]
	\caption{Data-driven synthesis of KKL observer given snapshots}\label{alg:3}
	\KwData{$\BRA{\bra{x^{(i)}, f^{-1}\bra{x^{(i)}} }: i\in \mbb{Z}_{1, n} }$}
	\KwResult{Mapping $\widehat{\zeta^\dagger}$}
	Set observer order $m$ and parameter $\beta\in (0,1)$\; 
	Set the $z$-kernel $\kappa$ and the $x$-kernel $\varkappa$\; 
	Compute kernel matrices $G_\vk, A_\vk, R_\vk$\;
	Choose mesh grid $\{\lambda_j\}_{j=1}^p \subset \partial\mbb{D}$ and threshold $\varepsilon_{\ms{res}}$\;
	\For{$j\in \mbb{Z}_{1,p}$}{
		Solve the eigenvalue problem $\eqref{eq:eigenvalue.problem}$ to determine $\hat{\psi}_j$\;
		Calculate the residual $\mr{res}(\lambda_j)$\;
		\If{$\ms{res}(\hat\psi_j, \lambda_j)>\varepsilon_{\ms{res}}$}{Discard the candidate point\;}
	}
	Sort the verified points and rename $\lambda_1, \dots, \lambda_p\in \partial\mbb{D}$\;
	Solve \eqref{eq:KRR.h} to determine $c^h_i$ in $\hat{h}$, $i\in\mbb{Z}_{1,n}$\;
	\For{$i\in \mbb{Z}_{1,n}$}{
		Calculate $\tilde{z}^{(i)} = \tilde{\zeta}\bra{x^{(i)}}$ according to \eqref{eq:zeta.sample.truncation.3}\;
	}
	\For{$k\in \mbb{Z}_{1,d_x}$}{
		Calculate $c_k$ by \eqref{eq:KRR.optimization} with sample $\BRA{x^{(i)}, \tilde{z}^{(i)}}_{i=1}^n$\;
		Let $\widehat{\zeta_k^\dagger}$ be given as in \eqref{eq:KRR.function.form}\;
	}
\end{algorithm}   

\begin{remark}
    Minimizing the $\mc{L}^2$ residual in problem \eqref{eq:residual.minimization.L2} would require the quadrature $\int_{\mbb{X}} \vk(x^{(i)}, x) \vk(x^{(j)}, x)\xD{x}$. To avoid it, residual dynamic mode decomposition (ResDMD) \cite{colbrook2024rigorous} can be adopted, where the RKHS metric is used instead, i.e.,
    $$\ms{res}(\lambda_j) = \min_{0\neq \hat{\psi}_j \in \mc{N}_\vk } \widehat{\ms{res}}(\lambda_j, \hat{\psi}_j) = \min_{v_j\neq 0} \frac{\|((K^\ast)^{-1} - \lambda_j)\hat{\psi}_j\|_{\mc{N}_\vk}}{ \|\hat{\psi}_j\|_{\mc{N}_\vk}}$$ 
    is minimized. Coefficients $v_j$ can be found by solving an eigenvalue problem:
    \begin{equation}\label{eq:eigenvalue.problem}
        \min_{v_j^\ast G_\vk v_j=1} v_j^\ast \bra{R_\vk - \lambda_j A_\vk^\ast - \lambda_j^\ast A_\vk + |\lambda_j|^2 G_\vk} v_j.
    \end{equation}
    The $(i, i')$-th entry of $G_\vk$, $A_\vk$, and $R_\vk$ are $\vk\bra{x^{(i)}, x^{(i')}}$, $\vk\bra{f^{-1}\bra{x^{(i)}}, x^{(i')}}$, and $\vk\bra{f^{-1}\bra{x^{(i)}}, f^{-1}\bra{x^{(i')}}}$, respectively. 
    The theoretical analysis needs to replace $\delta_{\ms{mesh}}$ with $\delta_{\ms{mesh}}r_n$ in Lemma \ref{lem:well-spanning}, where $r_n$ is the metric distortion ratio on the $n$-dimensional subspace $\mc{G}$, specifically: $\max_{0\neq g\in \mc{G}_n} \|g\|_{\mc N_\kappa}/\|g\|_{\mc L^2}$. 
\end{remark}

\section{Numerical Experiments}\label{sec:numerical}
This section aims to demonstrate the numerical performance of the proposed three algorithms, using a classical Lorenz system that has a chaotic behavior:
$$\dot{x}_1 = 10(x_2-x_1), \, \dot{x}_2 = x_1(28-x_3)-x_2, \, \dot{x}_3 = x_1x_2-\frac{8}{3}x_3.$$
A sampling time of $0.01$ is used to convert the dynamics into a discrete-time one. We consider the state observation with measurement $y=h(x)=x_2$. To obtain each sample point on the Lorenz attractor, we randomly select a point on $[-15, 15]^3$ under a uniform distribution, and simulate the system forward for $3$ continuous time units (namely $300$ discrete time steps) in order to take the final state as the sample point. The observer is always chosen to be third-order ($m=3$).

\subsection{Examination of Algorithm \ref{alg:1}}
\begin{figure}[!t]
    \centering \includegraphics[width=0.65\columnwidth]{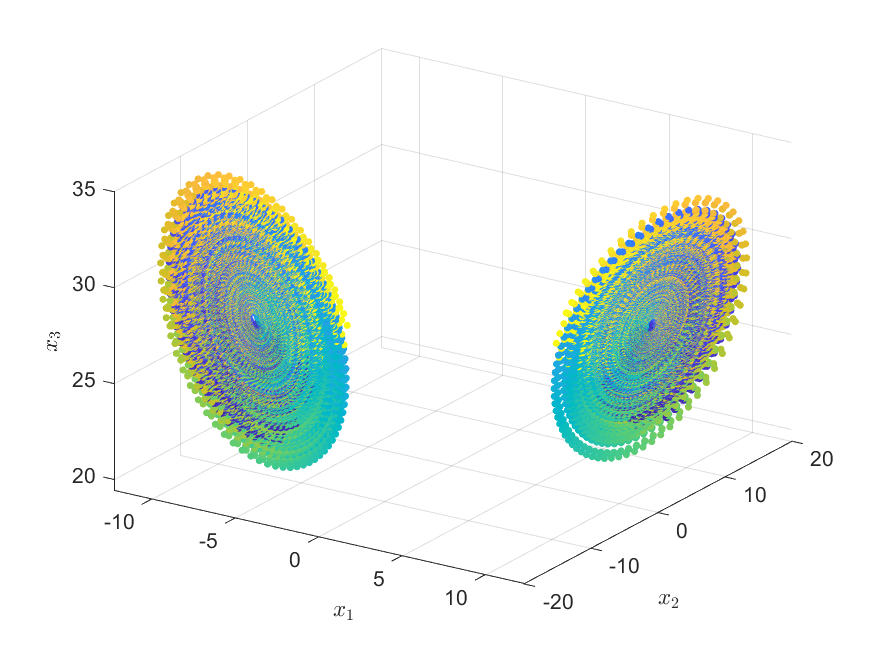}  
    \caption{Sample orbits from the Lorenz system.} 
    \label{fig:Lorenz_sample}
\end{figure} 
By first examining Algorithm \ref{alg:1}, we aim to verify the performance of the KKL observer affected by its truncation error and the KRR-based approximation of $\zeta^\dagger$ mapping. 
We first obtain $n=1000$ sample points on the attractor, whose distribution is visualized in Fig. \ref{fig:Lorenz_sample}. The color from yellow to blue indicates the time index from $\ell=-100$ to $\ell=0$ for every sample orbit. The collected orbits are located on the two sheaves of the Lorenz attractor, since the stay on the link between the two sheaves only takes very short time. 
\begin{figure}[!t]
    \includegraphics[width=\columnwidth]{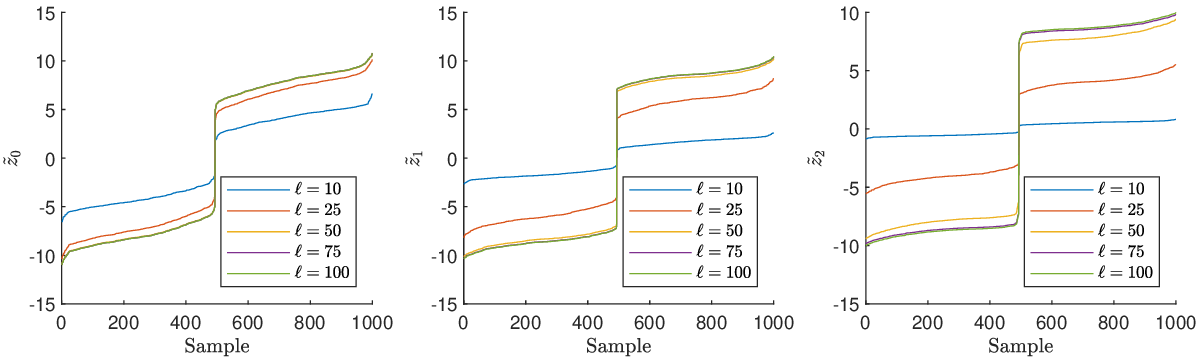}  
    \caption{Effect of truncation length on the approximation of $\tilde{z}$ value.} 
    \label{fig:Lorenz_tildez} 
\end{figure} 
Choosing $\beta=0.9$ and vary the truncation length $\ell$ from $10$ to $100$ (as a long enough orbit length, with $\beta^{100} < 3\times 10^{-5}$), the approximated values of $\tilde{\zeta}$ at the sample points are sorted and plotted as in Fig. \ref{fig:Lorenz_tildez}. It appears that $\ell=50$ (namely $0.5$ continuous time units) gives a good estimation. We will use this value of $\ell$ throughout the following investigations. 

\par
\begin{figure}[!t]
	\begin{center}
		\includegraphics[width=\columnwidth]{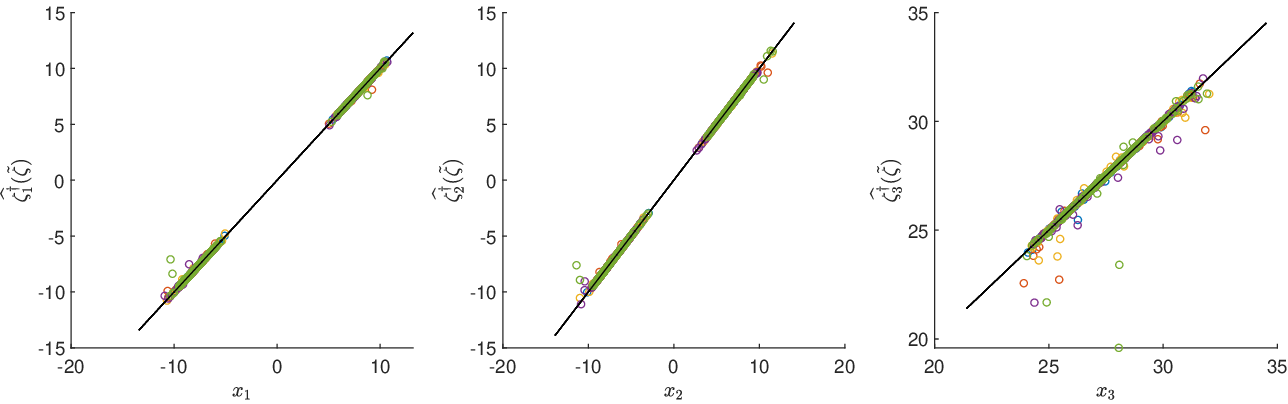}  
		\caption{Cross-validation of KRR for $\widehat{\zeta^\dagger}$ from many-orbit data. (Each color represent a ``fold'' in cross-validation. The black line is the $45^\circ$ line.)} 
		\label{fig:Lorenz_KRR_1} 
	\end{center}
\end{figure}

\begin{figure}[!t]
	\begin{center}
		\includegraphics[width=0.6\columnwidth]{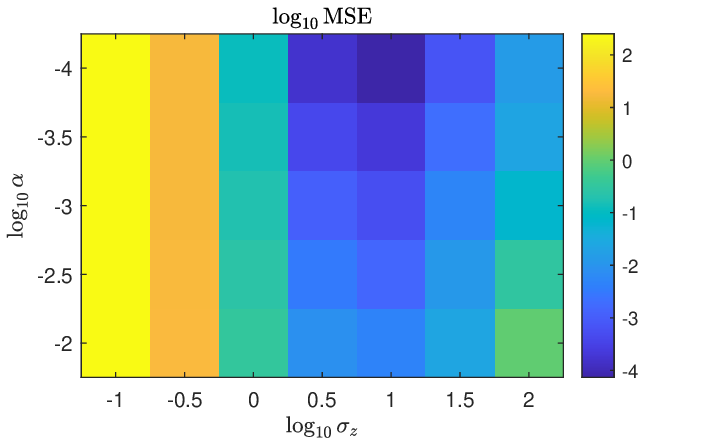}  
		\caption{Grid search for the hyperparameter tuning in KRR for $\widehat{\zeta^\dagger}$.} 
		\label{fig:Lorenz_KRR_cv} 
	\end{center}
\end{figure}

 We proceed with KRR, for which the kernel on the $z$-space is chosen as a Gaussian one $\kappa(z,z') = \exp(-\|z-z'\|^2/\sigma_z^2)$. In Fig. \ref{fig:Lorenz_KRR_1}, the actual $h(x)$ values of the validation data are plotted against the predicted values by trained model for $5$-fold cross-validation, under the choice of $\sigma_z=1$ and $\alpha=10^{-3}$. The mean-squared prediction error, $\|\widehat{\zeta^\dagger}(\tilde{z}) - x\|^2$, averaged among $5$ folds, is $0.1876$. 
For the fine-tuning of hyperparameters $\sigma_z$ and $\alpha$, we perform a grid search of these two parameters and look for a minimum of the mean-squared prediction error. The tuned value for $\sigma_z$ is $10$ and the regularization parameter is tuned to $\alpha = 10^{-4}$. 
The minimized cross-validated mean-squared prediction error is $7.37\times10^{-5}$. The grid search results are shown in Fig. \ref{fig:Lorenz_KRR_cv}. 
As such, the estimated pseudo-inverse $\zeta^\dagger$ has only a negligible loss. 

\subsection{Examination of Algorithm \ref{alg:2} and Algorithm \ref{alg:3}}
\begin{figure}[!t]
\centering
    \includegraphics[width=0.65\columnwidth]{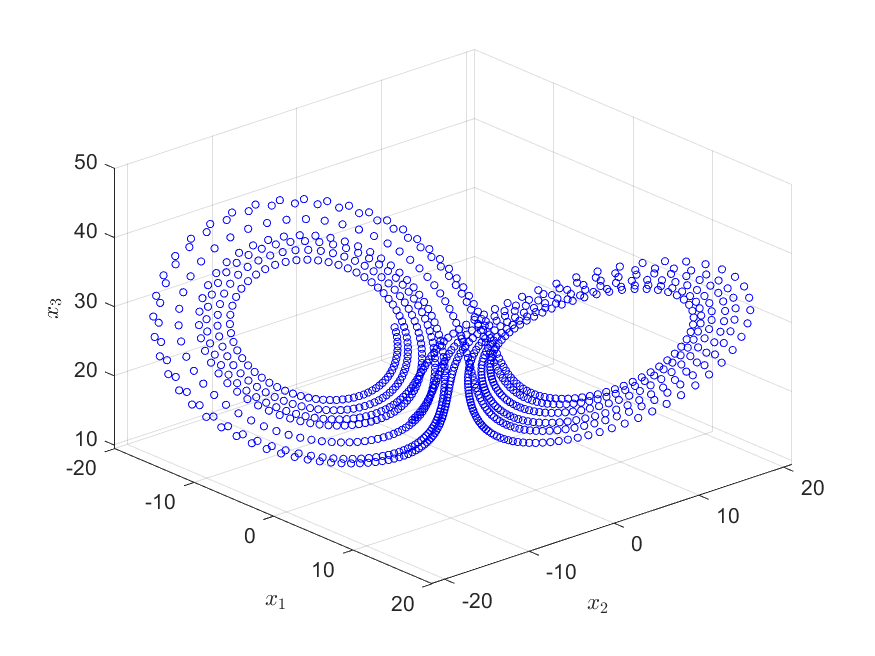}
    \caption{Single orbit data from the Lorenz system.} 
    \label{fig:Lorenz_single_orbit}
\end{figure}

Next by examining Algorithm \ref{alg:2}, we aim to verify that the numerical performance is retained when approximating the observer on a long orbit. For the data generation, we initialize the system at $(5, 5, 5)$ and take the data from $3$ to $13$ continuous time units. The data points are scattered on the attractor as shown in Fig. \ref{fig:Lorenz_single_orbit}. 

\begin{figure}[!t]
\centering
    \includegraphics[width=0.5\columnwidth]{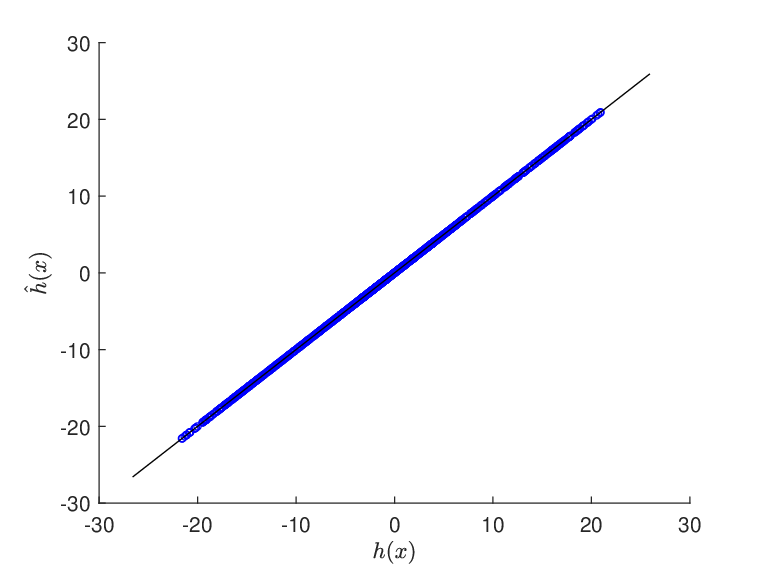}  
    \caption{Interpolation of $h$ by KRR.} 
    \label{fig:Lorenz_h_interpolate} 
\end{figure}

\begin{figure}[!t] 
\centering
    \includegraphics[width=\columnwidth]{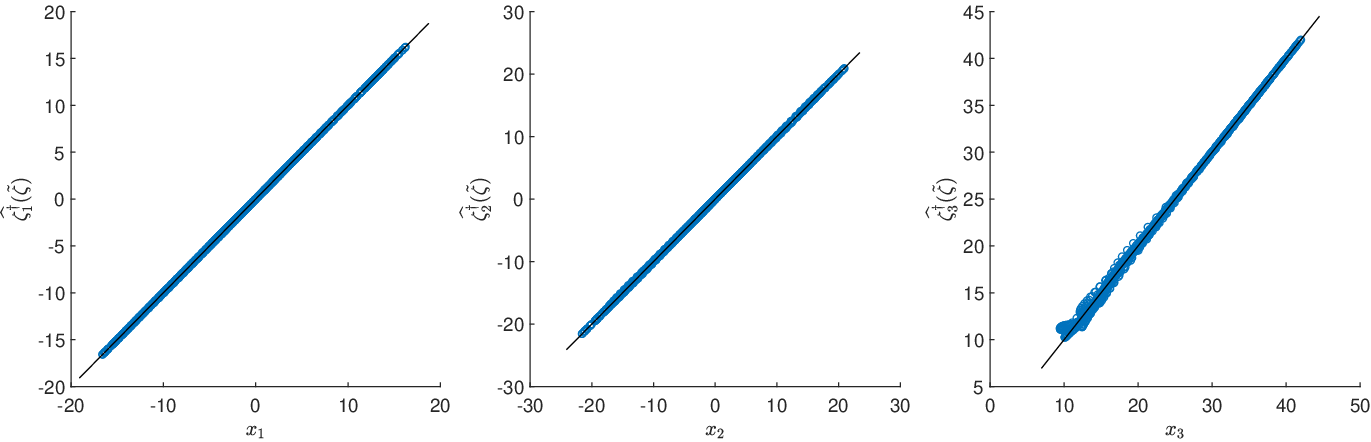}  
    \caption{The KRR for $\widehat{\zeta^\dagger}$ from single-orbit data.} 
    \label{fig:Lorenz_KRR_2}  
\end{figure}

The interpolation of $h$ by the kernel functions turns out to have very high precision, with a mean squared prediction error of $1.70\times 10^{-9}$ when using the Wendland kernel with $k=1$ and a bandwidth parameter $\sigma_x = 10$.
The interpolation result is visualized in Fig. \ref{fig:Lorenz_h_interpolate} against the $45^\circ$ line.
Again, performing KRR for the estimation of $\widehat{\zeta^\dagger}$ using the previously determined hyperparameters, we obtain a high-quality approximation of $x$, with a mean-squared error of $0.1111$. The comparison is shown in Fig. \ref{fig:Lorenz_KRR_2}. 

\begin{figure}[!t]
    \centering
    \includegraphics[width=0.65\columnwidth]{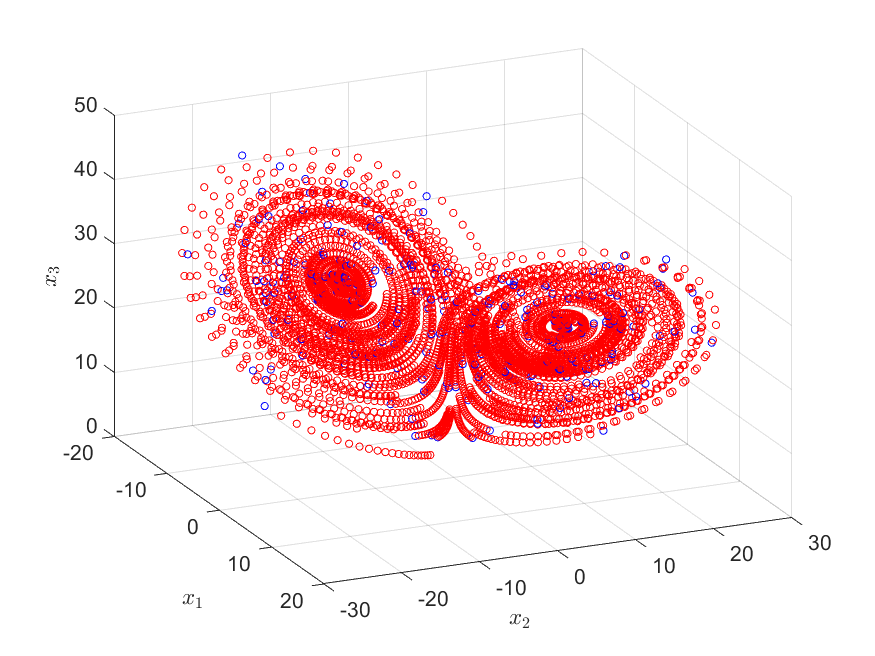}
    \caption{Snapshot data from the Lorenz system. Each red point is the successor of a blue point.} 
    \label{fig:Lorenz_snapshots}
\end{figure}

For Algorithm \ref{alg:3}, we collect snapshots of $5000$ predecessor-successor pairs from $250$ orbits, each containing a record of $20$ discrete time steps after the states are settled on the attractor, as shown in Fig. \ref{fig:Lorenz_snapshots}. 
The same Wendland kernel is chosen on the $x$-space. Then, dividing the unit circle on the complex plane $\partial\mbb{D}$ into $p$ intervals and choosing their mid-points as candidate approximate eigenvalues. 
Varying the number of approximate eigenvalues $p$ from $100$ to $2000$, the corresponding MSEs are shown in Table \ref{tab:p}. 

\begin{figure}[!t]
	\begin{center}
		\includegraphics[width=\columnwidth]{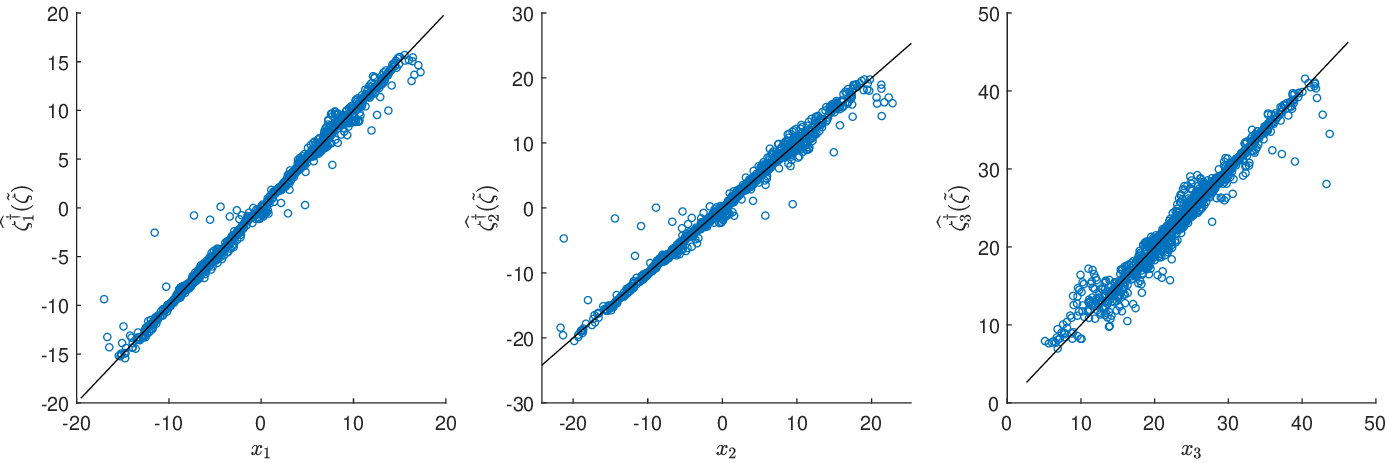}  
		\caption{The KRR for $\widehat{\zeta^\dagger}$ from snapshot data.} 
		\label{fig:Lorenz_KRR_3} 
	\end{center}
\end{figure}

\begin{table}[!t]
    \caption{MSE of state observation by approximate eigenfunctions.}\label{tab:p} 
    \centering
    \begin{tabular}{c|ccccccc}
        \toprule
        $p$ & $100$ & $200$ & $400$ & $600$ & $800$ & $1000$ & $2000$ \\
        \midrule 
        MSE& $53.55$ & $30.87$ & $17.05$ & $18.64$ & $4.39$ & $3.99$ & $4.07$ \\
        \bottomrule 
    \end{tabular}
\end{table}
We find that using $p=800$ gives a satisfactory KRR performance. 
The plot of the estimated $x$ against the actual $x$ values are shown in Fig. \ref{fig:Lorenz_KRR_3}.
Due to the need of extrapolating approximate eigenfunctions, the errors are more significant compared to the previous two algorithms. 

\par To examine if it is possible to reduce the number of approximate eigenfunctions, we adopt a residual threshold $\varepsilon_{\ms{res}}$ and discard those with residuals lower than the threshold. The maximum residual among the $800$ approximate eigenfunctions is $0.0455$. The effect of this hyperparameter, varied from $0.04$ to $0.005$ is shown in Table \ref{tab:eres}. It then appears that the threshold should not be too low to preserve a enough large basis. 
\begin{table}[!t]
\caption{Effect of residual threshold for approximate eigenfunctions on state observation.}\label{tab:eres} 
\centering
    \begin{tabular}{c|cccccc}
        \toprule
        $\varepsilon_{\ms{res}}$ & $0.04$ & $0.03$ & $0.02$ & $0.01$ & $0.0075$ & $0.005$ \\
        \midrule
        $p$ & $758$ & $644$ & $510$ & $268$ & $116$ & $24$ \\
        MSE & $4.39$ & $4.40$ & $4.39$ & $4.48$ & $10.09$ & $40.91$ \\
        \bottomrule
    \end{tabular}
\end{table}

\subsection{Comparison with Other Synthesis Approaches}
Finally we compare the Koopman-based data-driven observer from the above algorithms to a model-based Luenberger observer designs. The model-based observer is designed based on the Lipschitz continuity of the nonlinear part of the model -- the method was proposed in \cite{rajamani2002observers}. We use a Lipschitz constant estimate of $1$ for the nonlinear terms. Clearly a Lipschitz constant of $1$ is an underestimate. However, the model-based synthesis is infeasible if a realistically high estimate, e.g., $50$ is used, while estimating the Lipschitz constant as $1$ still gives a practically well-performing observer. 
This demonstrates the shortage of a reliable model-based synthesis method for a chaotic system. We also attempted to compare with the extended Kalman filter (EKF), which in this example turned out to be numerically unstable. Such a phenomena should be owed to the severe nonlinearity of the system. 

\par
\begin{figure}[!t]
	\begin{center}
		\includegraphics[width=\columnwidth]{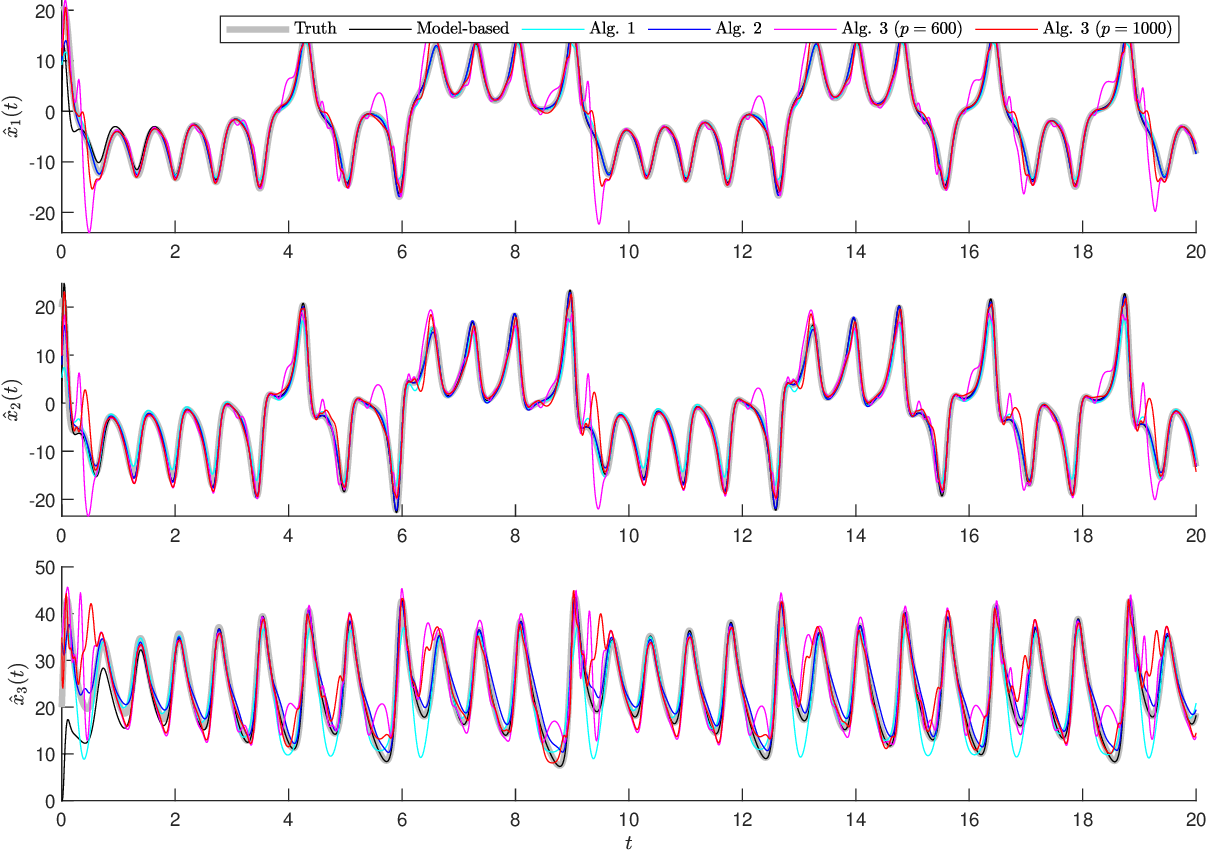}  
		\caption{Comparison of data-driven and model-based observers.} 
		\label{fig:Lorenz_observer} 
	\end{center} 
\end{figure}

 The comparison of the data-driven and the model-based observers using Lipschitz constant are shown in Fig. \ref{fig:Lorenz_observer} against the actual trajectories. 
At this point, we conclude that the data-driven observers achieve acceptable low errors; specifically, the mean-squared error of observation (taking the average after $t=3$) is $12.66$ (Algorithm \ref{alg:1}), $3.44$ (Algorithm \ref{alg:2}), $31.73$ (Algorithm \ref{alg:3} with $p=600$), and $17.31$ (Algorithm \ref{alg:3} with $p=1000$), respectively, vis-{\`{a}}-vis a variance of $218.81$ in the actual states $x$. 

\section{Conclusions}\label{sec:conclusions}
\par In this paper, data-driven, convex learning-based observer synthesis approaches for measure-preserving systems are proposed. 
By connecting the KKL observer's injective mapping and the Koopman operator, three algorithms are provided, based on various forms of available datasets -- many orbits, a single long orbit, or snapshots only. 
In the last difficult case, where no backtrackable history data is available, we need to adopt a kernel-based computation of the pseudo-eigenfunctions Koopman operator. 
Both theoretical proofs for the uniform bounds on the generalized observation error and numerical studies on a chaotic system have demonstrated the performance of the proposed data-driven synthesis. 

\par The current work considers only autonomous systems. From a control point-of-view, it would be of interest to consider controlled systems with measure-preserving drift terms and alongside control inputs.  
Furthermore, it is highly desirable to have a \emph{unified} approach for data-driven observer synthesis for nonlinear systems with convergent dynamics, divergent dynamics, measure-preserving dynamics, or mixed behaviors. It would be beneficial to develop a general understanding of the capacity of state observation, through deriving a ``universal'' Koopman-based observer.

\bibliographystyle{ieeetr}
\bibliography{mybib.bib}

\end{document}